\documentclass[letterpaper, 10 pt, conference]{ieeeconf}
\IEEEoverridecommandlockouts 

\usepackage{microtype}

\usepackage[pdftex,pdfauthor={Yu Fan Chen, Miao Liu, Michael Everett, and Jonathan P.\ 	How},pdftitle={Decentralized Multiagent Collision Avoidance}]{hyperref}
\hypersetup{colorlinks,linkcolor={green!50!black},citecolor={green!50!black},urlcolor={blue!80!black}}
\makeatletter \let\NAT@parse\undefined \makeatother
\usepackage[sort,compress]{cite}
\usepackage{graphicx} 
\usepackage{amsfonts}
\usepackage{amsmath,soul}
\usepackage{color}
\usepackage[font=small]{subcaption}
\usepackage{balance}
\usepackage[font=small]{caption}
\usepackage[linesnumbered,ruled,vlined]{algorithm2e}

\DeclareMathOperator*{\argmin}{\arg\!\min}
\DeclareMathOperator*{\argmax}{\arg\!\max}

\title{\LARGE \bf Decentralized Non-communicating Multiagent Collision Avoidance\\ with Deep Reinforcement Learning}

\author{Yu Fan Chen, Miao Liu, Michael Everett, and Jonathan P.\
  How
  \thanks{Laboratory of Information and Decision Systems,
    Massachusetts Institute of Technology, 77 Massachusetts Ave.,
    Cambridge, MA, USA\newline {\tt\footnotesize \{chenyuf2, miaoliu, mfe, jhow\}@mit.edu}}%
}

\usepackage[svgnames]{xcolor} \definecolor{DarkGreen}{rgb}{0,0.5,0}
\definecolor{DarkRed}{rgb}{0.75,0,0}

\usepackage{tikz,mathtools}
\usepackage[capitalize]{cleveref}
\crefformat{equation}{(#2#1#3)}
\Crefformat{equation}{Equation~(#2#1#3)}
\Crefname{equation}{Equation}{Equations}

\usetikzlibrary{shapes,positioning,automata,arrows,fit,backgrounds,calc}
\tikzstyle{block} = [draw, fill=blue!20, rectangle,minimum height=1em,
minimum width=2em] \tikzstyle{sum} = [draw, fill=blue!20, circle, node
distance=1cm] \tikzstyle{input} = [coordinate] \tikzstyle{output} =
[coordinate] \tikzstyle{pinstyle} = [pin edge={to-,thin,black}]
\usetikzlibrary{trees} \usetikzlibrary{decorations.pathmorphing}
\usetikzlibrary{decorations.markings}
\definecolor{darkgreen}{rgb}{0,0.5,0}
\definecolor{darkred}{rgb}{220,20,60}

\makeatletter
\renewcommand\paragraph{\@startsection{subsubsection}{4}{\z@}%
{0.25ex \@plus.5ex \@minus.2ex}%
{-.15em}%
{\normalfont\normalsize\itshape}}
\makeatother

\begin{document}

\maketitle
\thispagestyle{empty} \pagestyle{empty}

\begin{abstract} 
Finding feasible, collision-free paths for multiagent systems can be challenging, particularly in non-communicating scenarios where each agent's intent (e.g. goal) is unobservable to the others. In particular, finding time efficient paths often requires anticipating interaction with neighboring agents, the process of which can be computationally prohibitive. This work presents a decentralized multiagent collision avoidance algorithm based on a novel application of deep reinforcement learning, which effectively offloads the online computation (for predicting interaction patterns) to an offline learning procedure. Specifically, the proposed approach develops a value network that encodes the estimated time to the goal given an agent's joint configuration (positions and velocities) with its neighbors. Use of the value network not only admits efficient (i.e.,~real-time implementable) queries for finding a collision-free velocity vector, but also considers the uncertainty in the other agents' motion. 
Simulation results show more than 26\% improvement in paths quality (i.e., time to reach the goal) when compared with optimal reciprocal collision avoidance (ORCA), a state-of-the-art collision avoidance strategy.
\end{abstract}

\section{Introduction} \label{sec:intro}
Collision avoidance is central to many robotics applications, such as multiagent coordination~\cite{mellinger_mixed-integer_2012}, autonomous navigation through human crowds~\cite{choi_real-time_2014}, pedestrian motion prediction~\cite{kim_brvo:_2015}, and computer crowd simulation~\cite{guy_clearpath:_2009}. Yet, finding collision-free, time efficient paths around other agents remains challenging, because it may require predicting other agents' motion and anticipating interaction patterns, through a process that needs to be computationally tractable for real-time implementation. 

If there is a reliable communication network for agents to broadcast their intents (e.g. goals, planned paths), then collision avoidance can be enforced through a centralized planner. For instance, collision avoidance requirements can be formulated as separation constraints in an optimization framework for finding a set of jointly feasible and collision-free paths~\cite{mellinger_mixed-integer_2012,augugliaro_generation_2012,chen_decoupled_2015}. However, centralized path planning methods can be computationally prohibitive for large teams~\cite{mellinger_mixed-integer_2012}. To attain better scalability, researchers have also proposed distributed algorithms based on message-passing schemes~\cite{purwin_theory_2008,desaraju_decentralized_2011}, which resolve local (e.g. pairwise) conflicts without needing to form a joint optimization problem between all members of the team. 

This work focuses on scenarios where communication cannot be reliably established, which naturally arises when considering human-robot interactions and can also be caused by hardware constraints or failures. This limitation poses additional challenges for collision avoidance, because mobile agents would need to cooperate without necessarily having knowledge of the other agent's intents. Existing work on non-communicating collision avoidance can be broadly classified into two categories, reaction-based and trajectory-based. Reaction-based methods~\cite{snape_hybrid_2011,ferrer_social-aware_2013, berg_reciprocal_2011} specify one-step interaction rules for the current geometric configuration. For example, reciprocal velocity obstacle (RVO)~\cite{van_den_berg_reciprocal_2008} is a reaction-based method that adjusts each agent's velocity vector to ensure collision-free navigation. However, since reaction-based methods do not consider evolution of the neighboring agents' future states, they are short-sighted in time and have been found to create oscillatory and unnatural behaviors in certain situations~\cite{trautman_unfreezing_2010,ferrer_social-aware_2013}.

\begin{figure}[t]
	\centering
	\includegraphics [trim=0 60 0 0, clip, angle=0, width=.8\columnwidth, keepaspectratio]{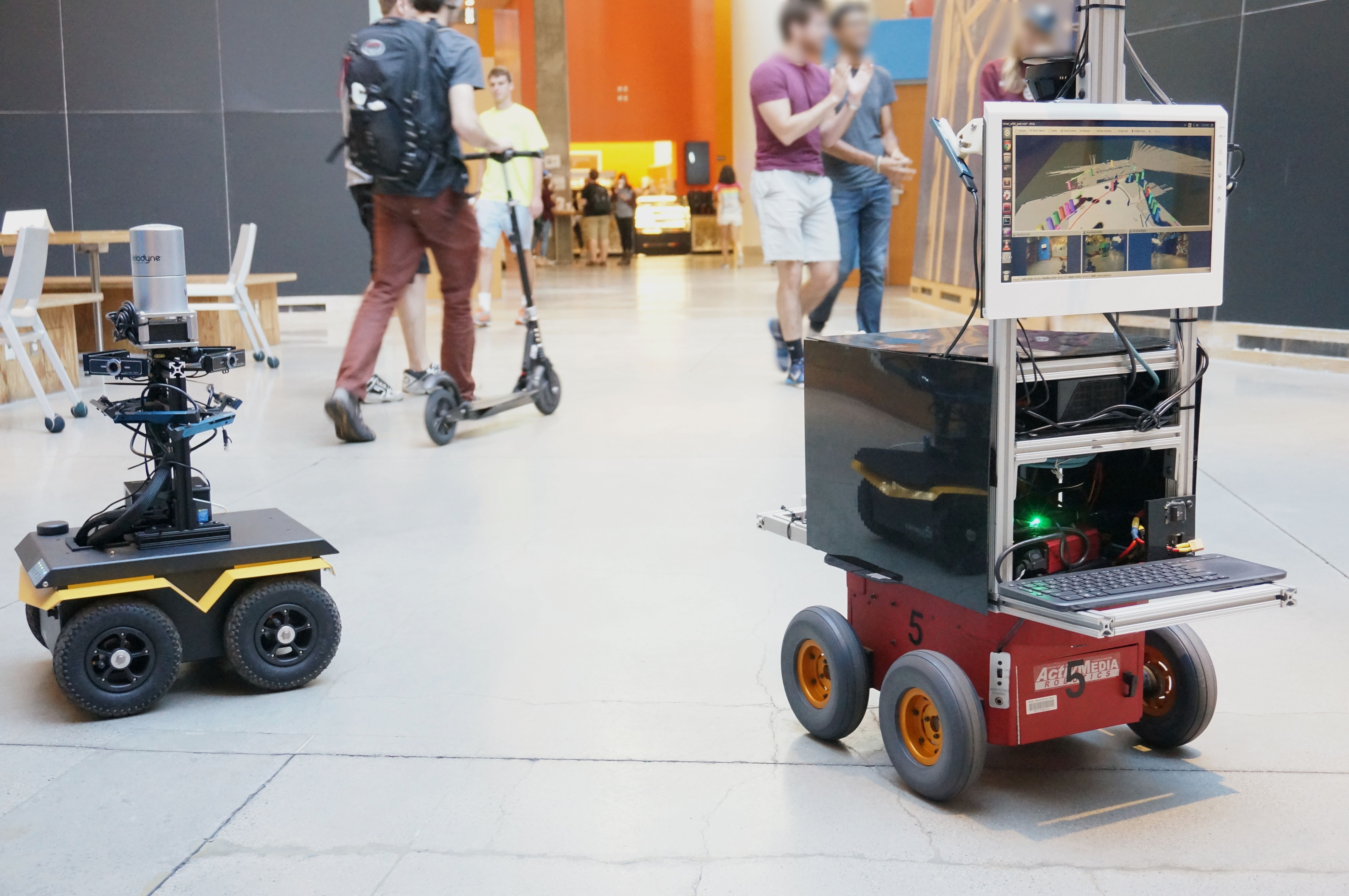}
	\caption{Autonomous ground vehicles navigating alongside pedestrians. Collision avoidance is essential for coordinating multiagent systems and modeling interactions between mobile agents.} 
	\label{fig:rover} 
	\vskip -0.1in
\end{figure}

In contrast, trajectory-based methods explicitly account for evolution of the joint (agent and neighbors) future states by anticipating other agents' motion. A subclass of non-cooperative approaches~\cite{phillips_sipp:_2011,aoude_probabilistically_2013} propagates the other agents' dynamics forward in time, and then plans a collision-free path with respect to the other agents' predicted paths. However, in crowded environments, the set of predicted paths often marks a large portion of the space untraversable/unsafe, which leads to the \emph{freezing robot problem}~\cite{trautman_unfreezing_2010}. A key to resolving this issue is to account for interactions, such that each agent's motion can affect one another. Thereby, a subclass of cooperative approaches~\cite{kretzschmar_socially_2016,trautman_robot_2013,kuderer_feature-based_2012} has been proposed, which first infers the other agents' intents (e.g. goals), then plans a set of jointly feasible paths for all neighboring agents in the environment. Cooperative trajectory-based methods often produce paths with better quality (e.g. shorter time for all agents to reach their goal) than that of reaction-based methods~\cite{kretzschmar_socially_2016}. However, planning paths for all other agents is computationally expensive, and such cooperative approach typically requires more information than is readily available (e.g. other agent's intended goal). Moreover, due to model and measurement uncertainty, the other agents' actual paths might not conform to the planned/predicted paths, particularly beyond a few seconds into the future. Thus, trajectory-based methods also need to be run at a high (sensor update) rate, which exacerbates the computational problem.

The major difficulty in multiagent collision avoidance is that anticipating evolution of joint states (paths) is desirable but computationally prohibitive. This work seeks to address this issue through reinforcement learning -- to offload the expensive online computation to an offline training procedure. Specifically, this work develops a computationally efficient (i.e., real-time implementable) interaction rule by learning a value function that implicitly encodes cooperative behaviors. 

The main contributions of this work are (i) a two-agent collision avoidance algorithm based on a novel application of deep reinforcement learning, (ii) a principled way for generalizing to more ($n>2$) agents, (iii) an extended formulation to capture kinematic constraints, and (iv) simulation results that show significant improvement in solution quality compared with existing reaction-based methods.

\section{Problem Formulation}\label{sec:prob_formulation}

\subsection{Sequential Decision Making}
A non-communicating multiagent collision avoidance problem can be formulated as a partially-observable sequential decision making problem. Let $\mathbf{s}_t, \, \mathbf{a}_t$ denote an agent's state and action at time $t$. The agent's state vector can be divided into two parts, that is $\mathbf{s}_t = [\mathbf{s}_t^o, \, \mathbf{s}_t^h]$, where $\mathbf{s}_t^o$ denotes the observable part that can be measured by all other agents, and $\mathbf{s}_t^h$ denotes the hidden part that is only known to the agent itself. In this work, let position and velocity vectors in 2D be denoted by $\mathbf{p}$ and $\mathbf{v}$, respectively; let action be the agent's velocity, $\mathbf{a} = \mathbf{v}$; let the observable states be the agent's position, velocity, and radius (size), $\mathbf{s}^o = [p_x, \, p_y, \, v_x, \, v_y, \, r] \in \mathbb{R}^{5}$; and let the hidden states be the agent's intended goal position, preferred speed, and heading angle, $\mathbf{s}^h = [p_{gx}, \, p_{gy}, \, v_{pref}, \, \theta] \in \mathbb{R}^{4}$. 

The following presents a two-agent\footnote{This formulation can be generalized to multiagent ($n>2$) scenarios by replacing the other agent's state $\tilde{\mathbf{s}^o_t}$ with all other agents' states $\tilde{\mathbf{S}_t^o} = [\tilde{\mathbf{s}^o_{1,t}}, \ldots, \tilde{\mathbf{s}^o_{n,t}}]$, and expand \cref{eqn:con_collision} to include all pairwise collision constraints.} collision avoidance problem formulation, where an agent's own state and the other agent's state are denoted by $\mathbf{s}$ and $\tilde{\mathbf{s}}$, respectively. The objective is to minimize the expected time, $\mathbb{E}[t_g]$, of an agent to reach its goal by developing a policy, $\pi: \left( \mathbf{s}_{0:t}, \, \tilde{\mathbf{s}}^o_{0:t} \right) \mapsto \mathbf{a}_t$, that selects an action given the observed state trajectories, 
\begin{align}
\argmin_{\pi\left(\mathbf{s}, \, \tilde{\mathbf{s}}^{o}\right)} \quad &\mathbb{E}  \left[t_g | \mathbf{s}_0, \, \tilde{\mathbf{s}}^o_0,  \, \pi, \, \tilde{\pi} \right] \label{eqn:cost} \\ 
s.t. \quad & ||\mathbf{p}_t - \tilde{\mathbf{p}}_t||_2 \geq r + \tilde{r} \qquad \forall t
		\label{eqn:con_collision} \\ 
	 \quad & \mathbf{p}_{t_g} = \mathbf{p}_g \label{eqn:con_reach_goal} \\
	 \quad & \mathbf{p}_t = \mathbf{p}_{t-1} + \Delta t \cdot \pi( \mathbf{s}_{0:t}, \, \tilde{\mathbf{s}}^o_{0:t}) \nonumber \\
	 \quad & \tilde{\mathbf{p}}_t = \tilde{\mathbf{p}}_{t-1} + \Delta t \cdot \tilde{\pi}( \tilde{\mathbf{s}}_{0:t}, \, \mathbf{s}^o_{0:t}), 
	 	\label{eqn:con_kinematics}
\end{align}
where \cref{eqn:con_collision} is the collision avoidance constraint, \cref{eqn:con_reach_goal} is the goal constraint, \cref{eqn:con_kinematics} is the agents' kinematics, and the expectation in \cref{eqn:cost} is with respect to the other agent's policy and hidden states (intents). Note that static obstacles can be modeled as stationary agents, which will be discussed in more details in \cref{sec:results:non_coop}. 

Although it is difficult to solve for the optimal solution of \cref{eqn:cost}-\cref{eqn:con_kinematics}, this problem formulation can be useful for understanding the limitations of the existing methods. In particular, it provides insights into the approximations/assumptions made by existing works. A common assumption is reciprocity, that is $\pi=\tilde{\pi}$, such that each agent would follow the same policy~\cite{berg_reciprocal_2011,kretzschmar_socially_2016}. Thereby, the main difficulty is in handling the uncertainty in the other agent's hidden intents (e.g. goals). 

Reaction-based methods~\cite{berg_reciprocal_2011,ferrer_social-aware_2013} often specify a Markovian policy, $\pi( \mathbf{s}_{0:t}, \, \tilde{\mathbf{s}}^o_{0:t}) = \pi( \mathbf{s}_t, \, \tilde{\mathbf{s}}^o_t)$, that optimizes a one-step cost while satisfying collision avoidance constraints. For instance, in velocity obstacle approaches~\cite{berg_reciprocal_2011}, an agent chooses a collision-free velocity that is closest to its preferred velocity (i.e., directed toward its goal). Given this one-step nature, reaction-based methods do not anticipate the other agent's hidden intent, but rather rely on a fast update rate to react quickly to the other agent's motion. Although computationally efficient given these simplifications, reaction-based methods are myopic in time, which can sometimes lead to generating unnatural trajectories~\cite{trautman_robot_2013} (e.g.,~\cref{fig:training_a}).

Trajectory-based methods~\cite{kretzschmar_socially_2016,trautman_robot_2013,kuderer_feature-based_2012} solve \cref{eqn:cost}-\cref{eqn:con_kinematics} in two steps. First, the other agent's hidden state is inferred from its observed trajectory, $\hat{\tilde{\mathbf{s}}}^h_t = f(\tilde{\mathbf{s}}^o_{0:t})$, where $f(\cdot)$ is a inference function. Second, a centralized path planning algorithm, $\pi( \mathbf{s}_{0:t}, \, \tilde{\mathbf{s}}^o_{0:t}) = \pi_{central}( \mathbf{s}_t, \, \tilde{\mathbf{s}}^o_t, \, \hat{\tilde{\mathbf{s}}}^h_t)$, is employed to find jointly feasible paths. By planning/anticipating complete paths, trajectory-based methods are no longer myopic. However, both the inference and the planning steps are computationally expensive, and need to be carried out online at each new observation (sensor update $\tilde{\mathbf{s}}^o_t$).

Our approach uses a reinforcement learning framework to solve \cref{eqn:cost}-\cref{eqn:con_kinematics} by pre-computing a value function $V\left(\mathbf{s}, \, \tilde{\mathbf{s}}^{o}\right)$ that estimates the expected time to the goal. 
As a result, the proposed method offloads computation from the online planning step (as in trajectory-based methods) to an offline learning procedure. The learned value function enables the use of a computationally efficient one-step lookahead operation, which will be defined in \cref{eqn:multi} and explained in \cref{sec:approach}. 
Repeating this one-step lookahead operation at each sensor update leads to generating better paths, as shown later in \cref{fig:training_d}. 

\subsection{Reinforcement Learning} \label{sec:prob:RL}
Reinforcement learning (RL)~\cite{sutton_introduction_1998} is a class of machine learning methods for solving sequential decision making problems with unknown state-transition dynamics. Typically, a sequential decision making problem can be formulated as a Markov decision process (MDP), which is defined by a tuple $M=\langle S,A,P,R,\gamma\rangle$, where $S$ is the state space, $A$ is the action space, $P$ is the state-transition model, $R$ is the reward function, and $\gamma$ is a discount factor. By detailing each of these elements and relating to \cref{eqn:cost}-\cref{eqn:con_kinematics}, the following provides a RL formulation of the two-agent collision avoidance problem. 

\paragraph*{State space} The system's state is constructed by concatenating the two agents' individual states, $\mathbf{s}^{jn} = \left[ \mathbf{s}, \; \tilde{\mathbf{s}}^o \right] \in \mathbb{R}^{14}$. 

\paragraph*{Action space} The action space is the set of permissible velocity vectors. Here, it is assumed that an agent can travel in any direction at any time, that is $\mathbf{a}(\mathbf{s}) = \mathbf{v}$ for $||\mathbf{v}||_2 < v_{pref}$. It is also straightforward to impose kinematic constraints, which will be explored in \cref{sec:approach:rotate_constr}.
\begin{figure}[t]
	\centering
	\begin{subfigure}{0.235\textwidth}
	\centering
	\includegraphics [trim=0 0 695 0, clip, width=1.0\textwidth, keepaspectratio]{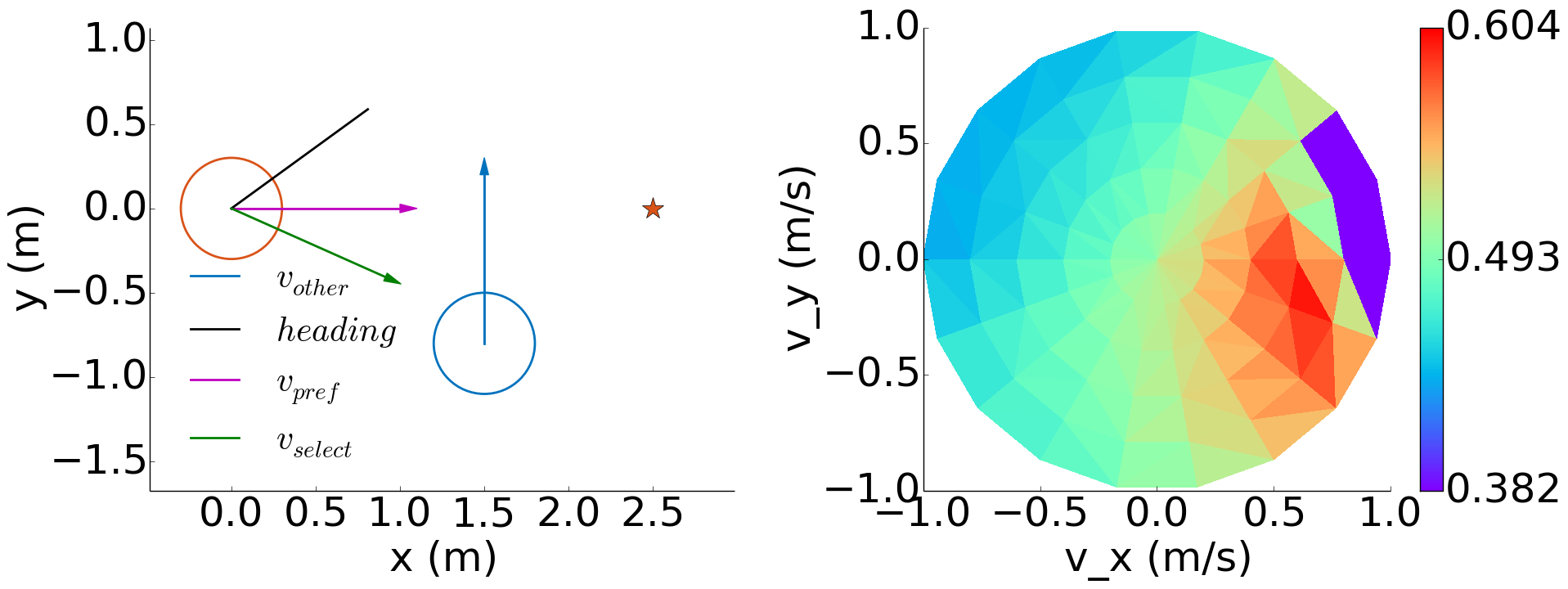}
	\caption{Input joint state}
	\label{fig:input_output_no_constr_a} 
	\end{subfigure}
	\begin{subfigure}{0.235\textwidth}
	\centering
	\includegraphics [trim=695 0 0 0, clip, width=1.0\textwidth, keepaspectratio]{figures/value_func_no_constr}
	\caption{Value function}
	\label{fig:input_output_no_constr_b} 
	\end{subfigure}
	\caption{RL policy. (a) shows a joint state of the system (geometric configuration) in the red agent's reference frame, with its goal aligned with the x-axis and marked as a star. (b) shows the red agent's value function at taking each possible action (velocity vector). Given the presence of the blue agent, ORCA~\cite{berg_reciprocal_2011} would choose an action close to the current heading angle (black vector), whereas the RL policy chooses to cut behind (green vector) the blue agent, leading to generating better paths similar to that in \cref{fig:training}.}
	\label{fig:input_output_no_constr} 
	\vskip -0.1in
\end{figure}

\paragraph*{Reward function}  A reward function is specified to award the agent for reaching its goal~\cref{eqn:con_reach_goal}, and penalize the agent for getting too close or colliding with the other agent~\cref{eqn:con_collision},
\begin{align}
R(\mathbf{s}^{jn}, \mathbf{a})  =  
	\begin{cases}
		-0.25 & \text{if} \quad d_{min} < 0 \\
		-0.1-d_{min}/2 & \text{else if}  \quad d_{min} <  0.2\\
		1 & \text{else if}  \quad \mathbf{p} = \mathbf{p}_g \\
		0 & \text{o.w.}
	\end{cases},
	\label{eqn:reward}
\end{align} 
where $d_{min}$ is the minimum separation distance between the two agents within a duration of $\Delta t$, assuming the agent travels at velocity $\mathbf{v}=\mathbf{a}$, and the other agent continues to travel at its observed velocity $\tilde{\mathbf{v}}$.
Note that the separation $d_{min}$ can be calculated analytically through simple geometry.

\paragraph*{State transition model}
\indent A probabilistic state transition model, $P( \mathbf{s}^{jn}_{t+1}, \mathbf{s}^{jn}_t|\mathbf{a}_t)$, is determined by the agents' kinematics as defined in~\cref{eqn:con_kinematics}. Since the other agent's choice of action also depends on its policy and hidden intents (e.g. goal), the system's state transition model is unknown. As in existing work~\cite{berg_reciprocal_2011}, this work also assumes reciprocity $\pi=\tilde{\pi}$, which leads to the interesting observation that the state transition model depends on the agent's learned policy.

\paragraph*{Value function} The objective is to find the optimal value function
\begin{align}
V^*(\mathbf{s}_0^{jn}) &= \sum_{t=0}^T \gamma^{t \cdot v_{pref}} \, R(\mathbf{s}^{jn}_t, \pi^*(\mathbf{s}^{jn}_t)),
\label{eqn:optimal_value}
\end{align}
where $\gamma\in[0,1)$ is a discount factor. Recall $v_{pref}$ is an agent's preferred speed and is typically time invariant. It is used here as a normalization factor for numerical reasons, because otherwise the value function of a slow moving agent could be very small. The optimal policy can be retrieved from the value function, that is 
\begin{align}
& \pi^*(\mathbf{s}^{jn}_0) = \argmax_{\mathbf{a}} R(\mathbf{s}_0, \mathbf{a}) + \nonumber \\ 
& \qquad \qquad \qquad \gamma^{\Delta t \cdot v_{pref}} \int_{\mathbf{s}_1^{jn}}P(\mathbf{s}^{jn}_0, \mathbf{s}^{jn}_{1}|\mathbf{a}) V^*(\mathbf{s}_1^{jn})d\mathbf{s}_1^{jn}. \label{eqn:optimal_policy}
\end{align} 
This work chooses to optimize $V(\mathbf{s}^{jn})$ rather than the more common choice $Q(\mathbf{s}^{jn}, \mathbf{a})$, because unlike previous works that focus on discrete, finite action spaces~\cite{mnih-dqn-2015,silver_mastering_2016}, the action space here is continuous and the set of permissible velocity vectors depends on the agent's state (preferred speed).

\section{Approach} \label{sec:approach}
The following presents an algorithm for solving the two-agent RL problem formulated in~\cref{sec:prob:RL}, and then generalizes its solution (policy) to multiagent collision avoidance. While applications of RL are typically limited to discrete, low-dimensional domains, recent advances in Deep RL~\cite{mnih-dqn-2015,silver_mastering_2016,zhang_learning_2016} have demonstrated human-level performance in complex, high-dimensional spaces. Since the joined state vector $\mathbf{s}^{jn}$ is in a continuous 14 dimensional space, and because a large amount of training data can be easily generated in a simulator, this work employs a fully connected deep neural network with ReLU nonlinearities to parametrize the value function, as shown in \cref{fig:convergence_a}. This value network is denoted by $V(\cdot;\mathbf{w})$, where $\mathbf{w}$ is the set of weights in the neural network. 

\subsection{Parametrization}
From a geometric perspective, there is some redundancy in the parameterization of the system's joint state $\mathbf{s}^{jn}$, because the optimal policy should be invariant to any coordinate transformation (rotation and translation). To remove this ambiguity, an agent-centric frame is defined, with the origin at the agent's position, and the x-axis pointing toward the agent's goal, that is, 
\begin{align}
	\mathbf{s}' &= \mathrm{\tt rotate}\left( \mathbf{s}^{jn} \right)  \nonumber \\
	& = [d_g, \; v_{pref}, \; v'_x, \; v'_y, \; r, \; \theta', \; \tilde{v}'_x, \; \tilde{v}'_y, \;
		\tilde{p}'_x, \; \tilde{p}'_y, \; \tilde{r}, \nonumber \\ 
	& \qquad	r+\tilde{r}, \; \cos(\theta'), \; \sin(\theta'), \; d_a ], \label{eqn:agent_centric}
\end{align} 
where $d_g=||\mathbf{p}_g - \mathbf{p}||_2$ is the agent's distance to goal, and $d_a=||\mathbf{p} - \tilde{\mathbf{p}}||_2$ is the distance to the other agent. An illustration of this parametrization is shown in \cref{fig:input_output_no_constr_a}. Note that this agent-centric parametrization is only used when querying the neural network.

\subsection{Generating Paths Using a Value Network}
Given a value network $V$, an RL agent can generate a path to its goal by repeatedly maximizing an one-step lookahead value~\cref{eqn:optimal_policy}, as outlined in~\cref{alg:CADRL}. This corresponds to choosing the action that on average, leads to the joint state with the highest value. However, the integral in~\cref{eqn:optimal_policy} is difficult to evaluate, because the other agent's next state $\tilde{\mathbf{s}}_{t+1}^o$ has an unknown distribution (depends on its unobservable intent). We approximate this integral by assuming that the other agent would be traveling at a filtered velocity for a short duration $\Delta t$ (line 5-6)\footnote{This work calculates the average velocity of the past 0.5 seconds and sets $\Delta t$ to 1.0 second.}. The use of a filtered velocity addresses a subtle oscillation problem as discussed in~\cite{van_den_berg_reciprocal_2008}. This propagation step amounts to predicting the other agent's motion with a simple linear model, which has been shown to produce good accuracy over small time scales~\cite{bera_realtime_2014}. It is important to point out that this approximation is \emph{not} assuming a linear motion model for $t > \Delta t$; uncertainty in the other agent's future motion is captured in the projected next state's value, $V(\hat{\mathbf{s}}_{t+1}, \, \hat{\tilde{\mathbf{s}}}^o_{t+1})$. Furthermore, the best action is chosen from a set of permissible\footnote{This work uses 25 pre-computed actions (e.g. directed toward an agent's goal or current heading) and 10 randomly sampled actions.} velocity vectors (line  8). An example of this one-step lookahead operation is visualized in~\cref{fig:input_output_no_constr_a}, in which the red agent chooses the green velocity vector to cut behind the blue agent, because this action maximizes the value of the projected state shown in~\cref{fig:input_output_no_constr_b}. 

\begin{algorithm}[t]
	\textbf{Input:} value network $\mathbf{V(\cdot;\mathbf{w})}$\\
	\textbf{Output:} trajectory $\mathbf{s}_{0:t_f}$ \\
	\While{not reached goal}{ 
		update $t$, receive new measurements $\mathbf{s}_t, \; \tilde{\mathbf{s}}^o_t$ \\
		$\hat{\tilde{\mathbf{v}}}_t$ $\leftarrow$ filter($\tilde{\mathbf{v}}_{0:t}$) \\
		$\hat{\tilde{s}}^o_{t+1}$ $\leftarrow$ propagate(${\tilde{\mathbf{s}}}^o_{t}, 
		\Delta t \cdot \hat{\tilde{\mathbf{v}}}_t $) \\
		$A$ $\leftarrow$ sampleActions() \\
		$\mathbf{a}_t $ $\leftarrow$ $\argmax_{\mathbf{a}_t \in A} R(\mathbf{s}^{jn}_t, \mathbf{a}_t) +  \bar{\gamma}V(\hat{\mathbf{s}}_{t+1},  \hat{\tilde{\mathbf{s}}}^o_{t+1})$
		\newline \hspace*{0.0cm} where  $\bar{\gamma} \leftarrow \gamma^{\Delta t \cdot v_{pref}}$, $\, \hat{\mathbf{s}}_{t+1}$ $\leftarrow$ propagate($\mathbf{s}_{t}, \Delta t \cdot \mathbf{a}_t $)
	}		
	\Return $\mathbf{s}_{0:t_f}$
	\caption{CADRL (Coll. Avoidance with Deep RL)}
	\label{alg:CADRL}
\end{algorithm} 

\begin{figure}[t]
	\centering
	\begin{subfigure}{0.18\textwidth}
		\centering
		\includegraphics [trim=195 105 520 200, clip, height=0.13\textheight, keepaspectratio]{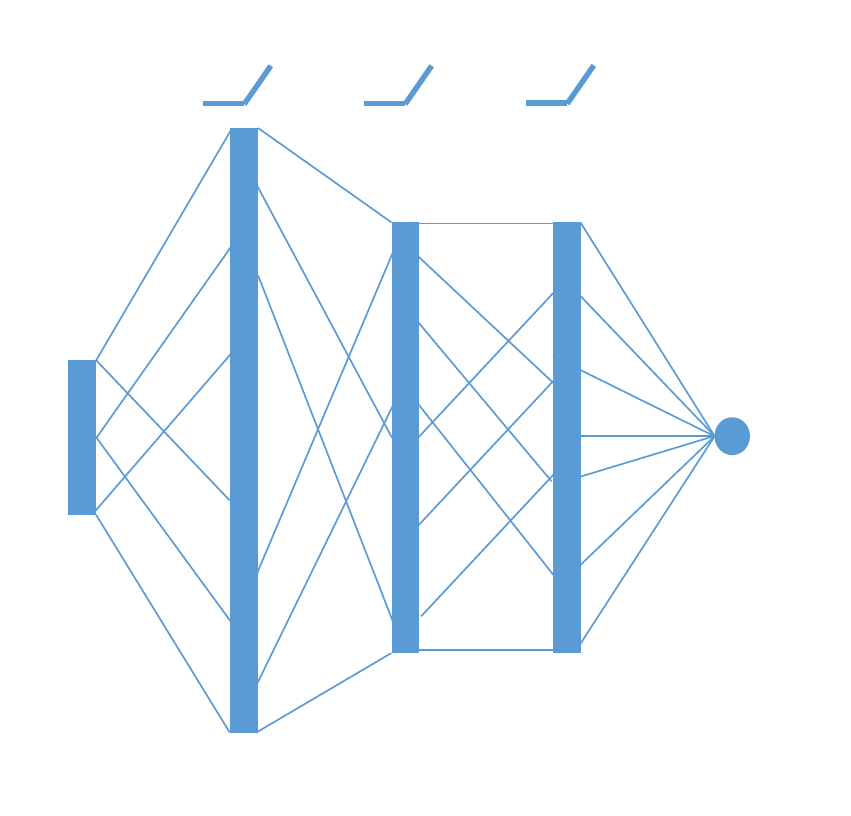}
		\caption{Value network}
		\label{fig:convergence_a} 
	\end{subfigure}
	\begin{subfigure}{0.28\textwidth}
		\centering
		\includegraphics [trim=0 0 0 0, clip, height=0.13\textheight, 		keepaspectratio]{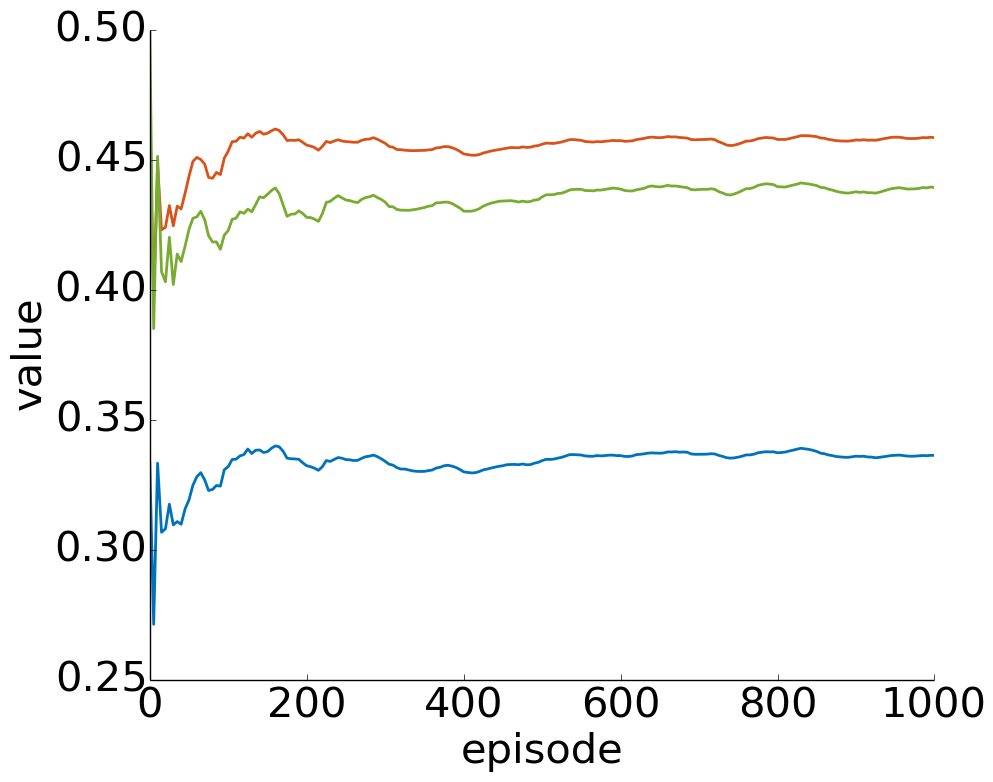}
		\caption{Convergence}
		\label{fig:convergence_b} 
	\end{subfigure}
	\caption{Convergence of a Deep RL policy. (a) shows a neural network used to parameterize the value function. (b) shows the values of three distinct test cases converge as a function of RL training episodes. For example, the blue line corresponds to the test case shown in the bottom row of~\cref{fig:training}.} 
	\label{fig:convergence} 
	\vskip -0.1in
\end{figure}

\begin{algorithm}[t]
	\textbf{Input:} trajectory training set $D$ \\
	\textbf{Output:} value network $V(\cdot;\mathbf{w})$\\
	$V(\cdot;\mathbf{w})$ $\leftarrow$ train\_nn($D$) \hskip 1.7cm //step 1: initialization \\
	duplicate value net $ V' \leftarrow V$            \hskip 0.95cm //step 2: RL  \\
	initialize experience set $E \leftarrow D$ \\
	\For{episode={$1,\ldots, N_{eps}$}}{ 
		\For {m times}{
			$\mathbf{s}_0, \mathbf{\tilde{s}}_0$ $\leftarrow$ randomTestcase() \\
			$\mathbf{s}_{0:t_f}$ $\leftarrow$ CADRL($V$), 
			$\mathbf{\tilde{s}}_{0:\tilde{t}_f} $ $\leftarrow$ CADRL($V$) \\
			$\mathbf{y}_{0:T}, \; \tilde{\mathbf{y}}_{0:\tilde{t}_f}$ $\leftarrow$ findValues($ V'$, $\mathbf{s}_{0:t_f}$, $\tilde{\mathbf{s}}_{0:\tilde{t}_f}$) \\
			$E$ $\leftarrow$ assimilate$\left( E, (\mathbf{y}, \mathbf{s}^{jn})_{0:t_f}, (\tilde{\mathbf{y}}, \tilde{\mathbf{s}}^{jn})_{0:\tilde{t}_f} \right)$ \\
		}
		$e$ $\leftarrow$ randSubset($E$) \\
		$\mathbf{w}$ $\leftarrow$ backprop($e$) \\
		\For{every $C$ episodes}{
			 Evaluate($V$), $V'$ $\leftarrow$ $V$
		}
	}		
	\Return $ V $
	\caption{Deep V-learning}
	\label{alg:V-learning}
\end{algorithm} 
\subsection{Training a Value Network}
The training procedure, outlined in \cref{alg:V-learning}, consists of two major steps. First, the value network is initialized by supervised training on a set of trajectories generated by a baseline policy (line~3). 

Specifically, each training trajectory is processed to generate a set of state-value pairs, $\{(\mathbf{s}^{jn}, \; y)_k\}_{k=1}^N$, where $y = \gamma ^{{t_g} \cdot {v_{pref}}}$ and $t_g$ is the time to reach goal. The value network is trained by back-propagation to minimize a quadratic regression error, $\argmin_\mathbf{w}\sum_{k=1}^N \left(y_k-V(\mathbf{s}^{jn}_k;\mathbf{w})\right)^2$. This work uses optimal reciprocal collision avoidance (ORCA)~\cite{berg_reciprocal_2011} to generate a training set of 500 trajectories, which contains approximately 20,000 state-value pairs.

\begin{figure*}[t]
	\centering
	\begin{subfigure}{0.24\textwidth}
		\centering
		\includegraphics [trim=0 0 0 20, clip, width=1.0 \textwidth, angle = 0]{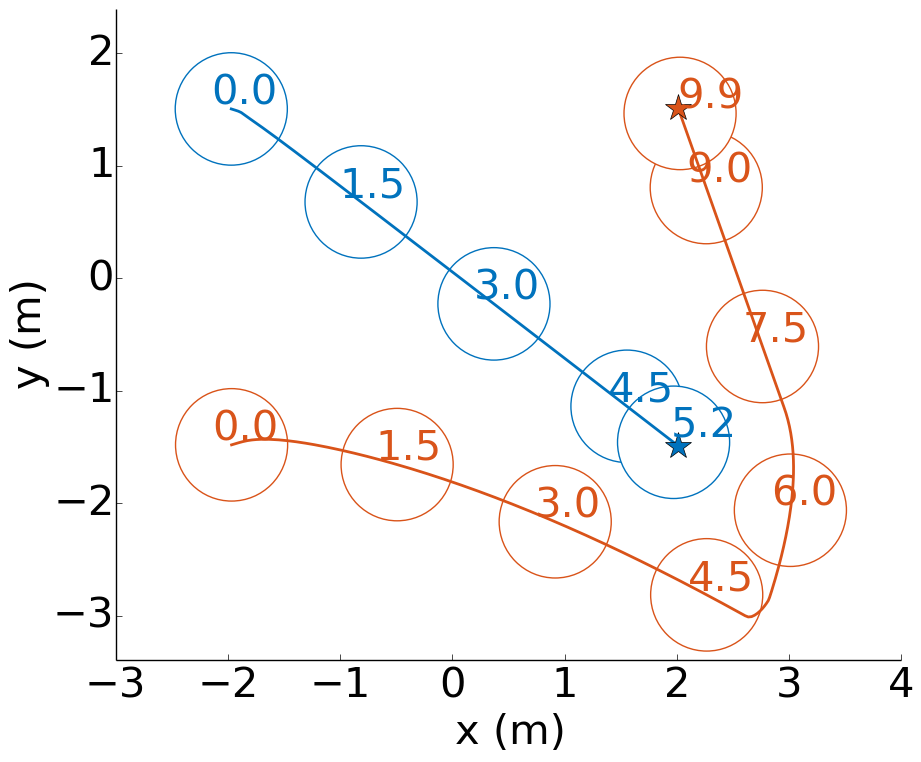}
		\includegraphics [trim=0 0 0 40, clip, width=1.0 \textwidth, angle = 0]{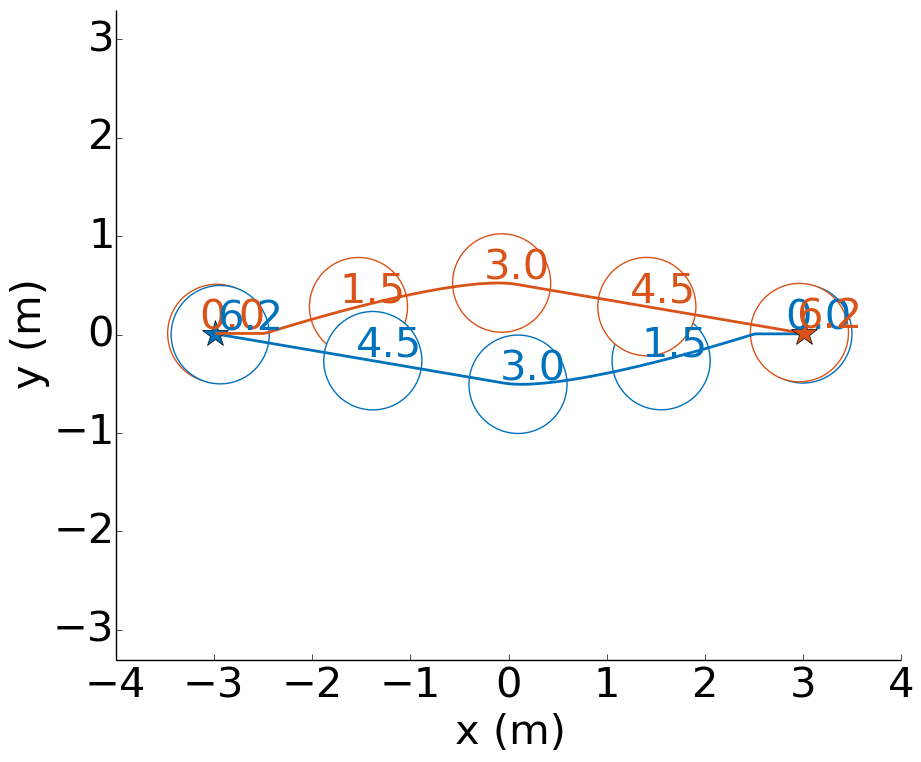}
		\caption{ORCA}
		\label{fig:training_a} 
	\end{subfigure}
	\begin{subfigure}{0.24\textwidth}
		\centering
		\includegraphics [trim=0 0 0 20, clip, width=1.0 \textwidth, angle = 0]{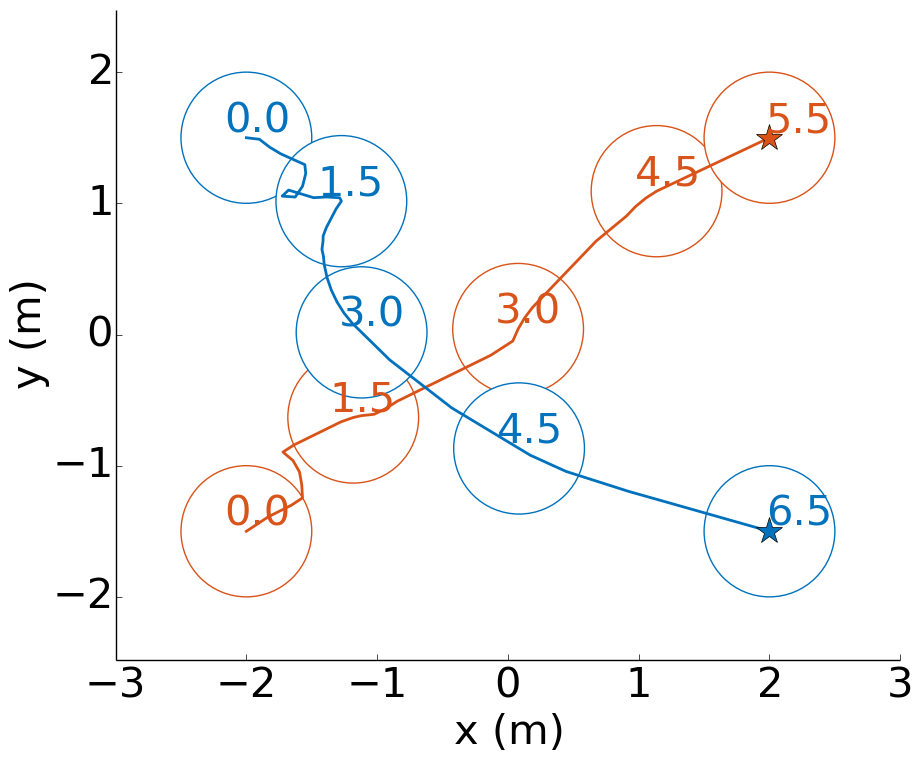}	
		\includegraphics [trim=0 0 0 40, clip, width=1.0 \textwidth, angle = 0]{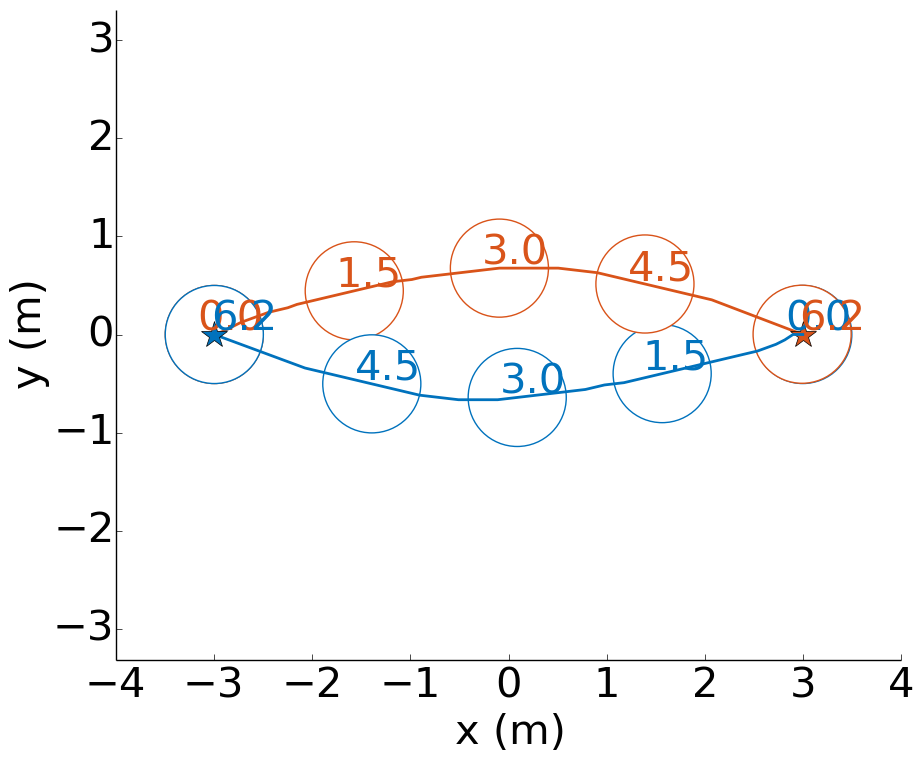}
		\caption{Episode 0}
		\label{fig:training_b} 
	\end{subfigure}
	\begin{subfigure}{0.24\textwidth}
		\centering
		\includegraphics [trim=0 0 0 20, clip, width=1.0 \textwidth, angle = 0]{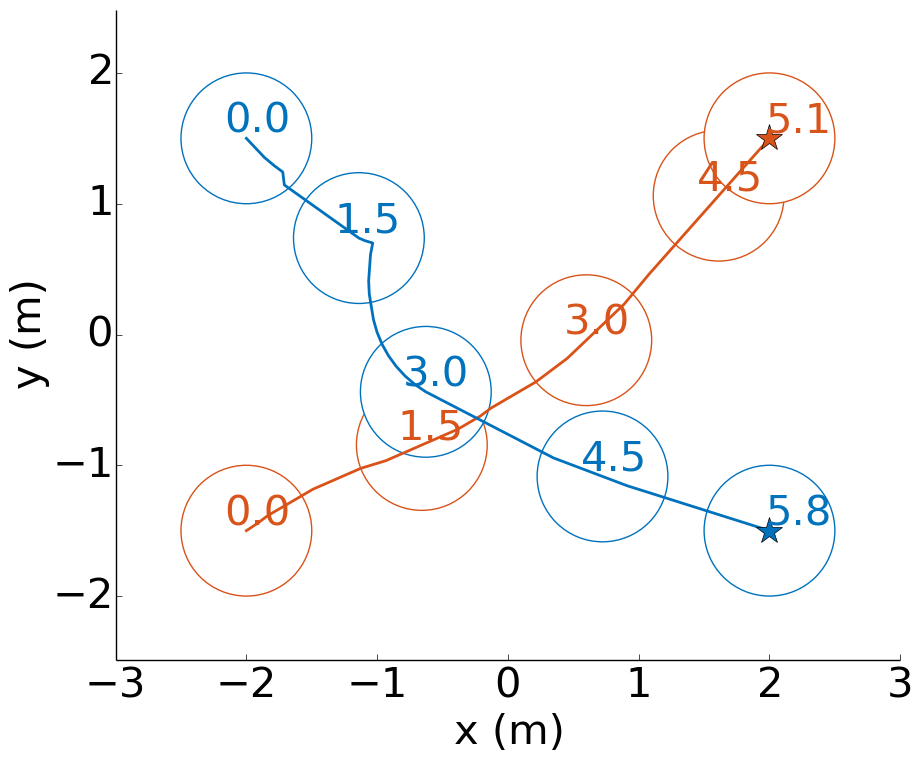}
		\includegraphics [trim=0 0 0 40, clip, width=1.0 \textwidth, angle = 0]{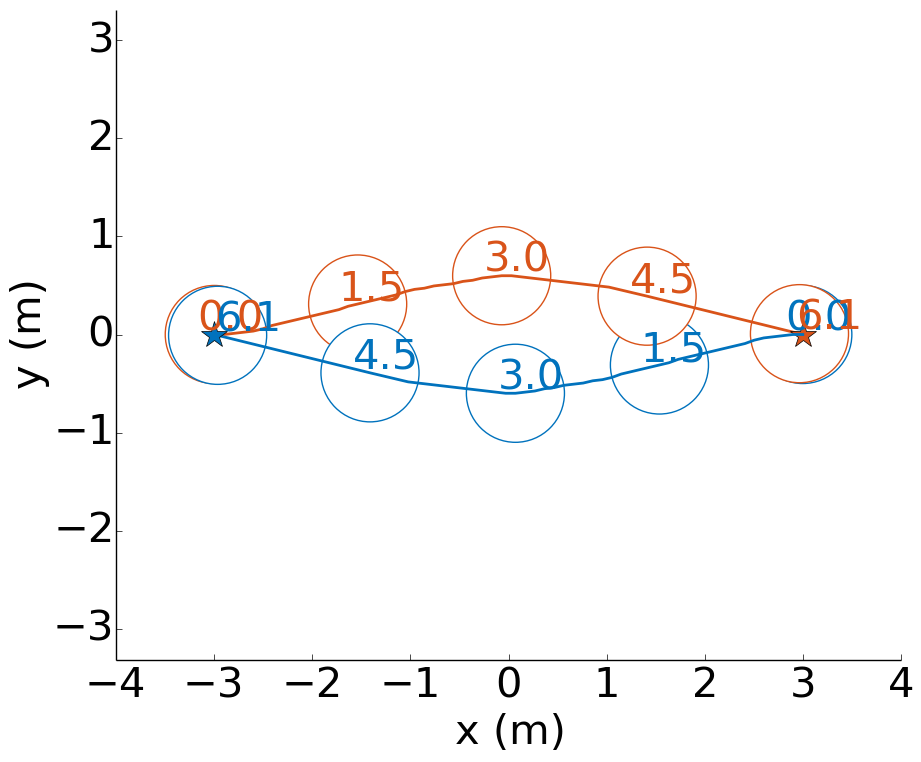}
		\caption{Episode 50}
		\label{fig:training_c} 
	\end{subfigure}
	\begin{subfigure}{0.24\textwidth}
		\centering
		\includegraphics [trim=0 0 0 20, clip, width=1.0 \textwidth, angle = 0]{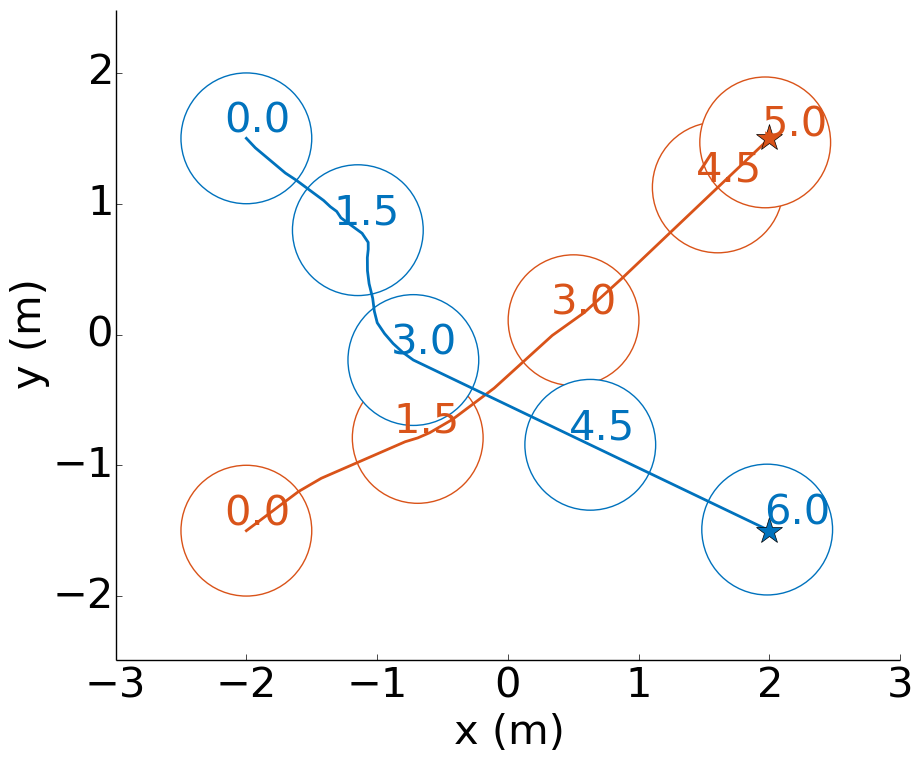}
		\includegraphics [trim=0 0 0 40, clip, width=1.0 \textwidth, angle = 0]{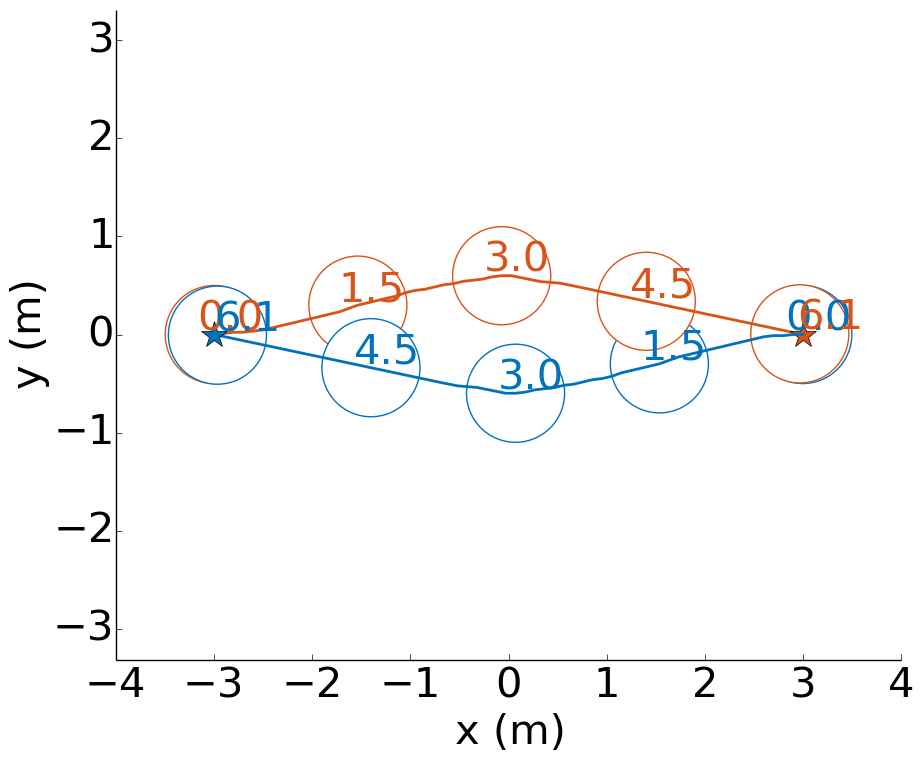}
		\caption{Episode 1000}
		\label{fig:training_d} 
	\end{subfigure}
	\caption{Training the value network. Circles show each agent's position at the labeled time, and stars mark the goals. (a) illustrates trajectories generated by the two agents each following ORCA~\cite{berg_reciprocal_2011}, and (b-d) illustrate trajectories generated by following the value network at various training episodes. Top (a) shows a test case which ORCA results in unnatural trajectories, where the red agent has traversed a large arc before reaching its goal. Top (b-d) show CADRL has learned to produce cooperative behaviors, as the blue agent slows down and cuts behind the red agent. Bottom (b-d) show the trajectories become more tight (better performing) during the training process, since the minimum separation $d_{min}$ reduces from 0.4m to 0.2m, as specified in \cref{eqn:reward}.}
	\label{fig:training} 
	\vskip -0.1in
\end{figure*}

We make a few remarks about this initialization step. First, the training trajectories do not have to be optimal. For instance, two of the training trajectories generated by ORCA~\cite{berg_reciprocal_2011} are shown in~\cref{fig:training_a}. The red agent was pushed away by the blue agent and followed a large arc before reaching its goal. Second, the initialization training step is not simply emulating the ORCA policy. Rather, it learns a time to goal estimate (value function), which can then be used to generate new trajectories following~\cref{alg:CADRL}. Evidently, this learned value function sometimes generates better (i.e. shorter time to goal) trajectories than ORCA, as shown in \cref{fig:training_b}. Third, this learned value function is likely to be suboptimal. For instance, the minimum separation $d_{min}$ between the two agents should be around 0.2m by \cref{eqn:reward}, but $d_{min}$ is greater than 0.4m (too much slack) in \cref{fig:training_b}.

The second training step refines the policy through reinforcement learning. Specifically, a few random test cases are generated in each episode (line 8), and two agents are simulated to navigate around each other using an $\epsilon$-greedy policy, which selects a random action with probability~$\epsilon$ and follows the value network greedily otherwise (line 9). The simulated trajectories are then processed to generate a set of state-value pairs (line 10). For convergence reasons, as explained in~\cite{mnih-dqn-2015}, rather than being used to update the value network immediately, the newly generated state-value pairs are assimilated (replacing older entries) into a large experience set $E$ (line 11). Then, a set of training points is randomly sampled from the experience set, which contains state-value pairs from many different simulated trajectories (line 12). The value network is finally updated by stochastic gradient descent (back-propagation) on the sampled subset. 

To monitor convergence of the training process, the value network is tested regularly on a few pre-defined evaluation test cases (line~15), two of which are shown in \cref{fig:training}. Note that the generated paths become tighter as a function of the number of training episodes (i.e. $d_{min}$ reduces from 0.4m to~0.2m). A plot of the test cases' values $V(\mathbf{s}^{jn}_0)$ 
shows that the value network has converged in approximately 800 episodes (\cref{fig:convergence_b}). It is important to point out that training/learning is performed on \emph{randomly} generated test cases (line 8), but not on the evaluation test cases.  

In addition to the standard Q-learning update~\cite{sutton_introduction_1998}, an important modification is introduced when calculating the state-value pairs (line 10). In particular, cooperation is encouraged by adding a penalty term based on a comparison of the two agents' extra time to reach the goal, $t_e = t_g - d_g / v_{pref}$. If $t_e < e_l$ and $\tilde{t}_e > e_u$\footnote{This work uses $e_l$ = 1.0 and $e_u$ = 2.0.}, which corresponds to a scenario where the agent reached its goal quickly but the other agent took a long time, an penalty of 0.1 would be subtracted from the training value. Albeit simple, this modification is crucial for discouraging aggressive behaviors such as exhibited by the blue agent in \cref{fig:training_a}. Without this modification, an agent would frequently travel straight toward its goal, expecting the other agent to yield.

\subsection{Incorporating Kinematic Constraints} \label{sec:approach:rotate_constr}
\begin{figure}[t]
	\centering
	\begin{subfigure}{0.1605\textwidth}
		\centering
		\includegraphics [trim=0 0 695 50, clip, width=1.0\textwidth, keepaspectratio]{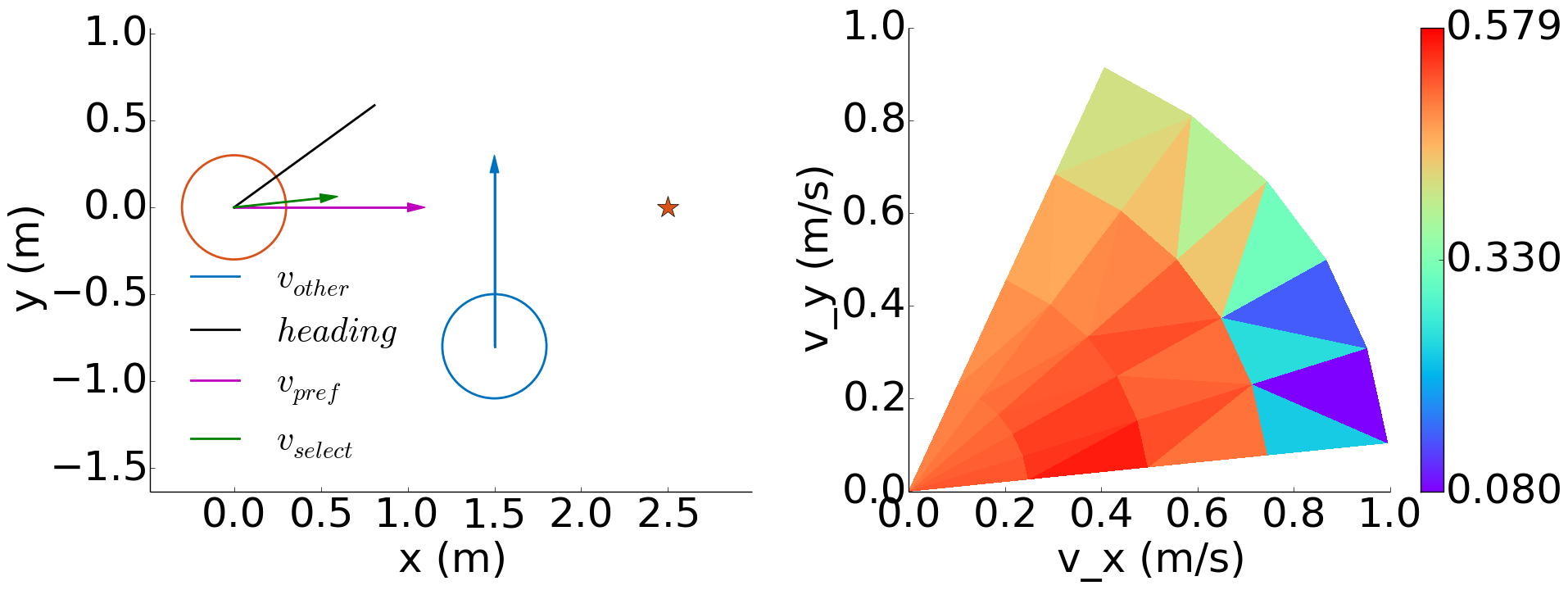}
		\includegraphics [trim=695 0 0 0, clip, width=1.0\textwidth, keepaspectratio]{figures/value_func_rotate_constr}
		\caption{Constrained action}
		\label{fig:rotate_constr_a} 
	\end{subfigure}
	\skip 0.01\textwidth
	\begin{subfigure}{0.29\textwidth}
		\centering
		\includegraphics [trim=0 0 0 0, clip, width=1.0\textwidth, 
			keepaspectratio]{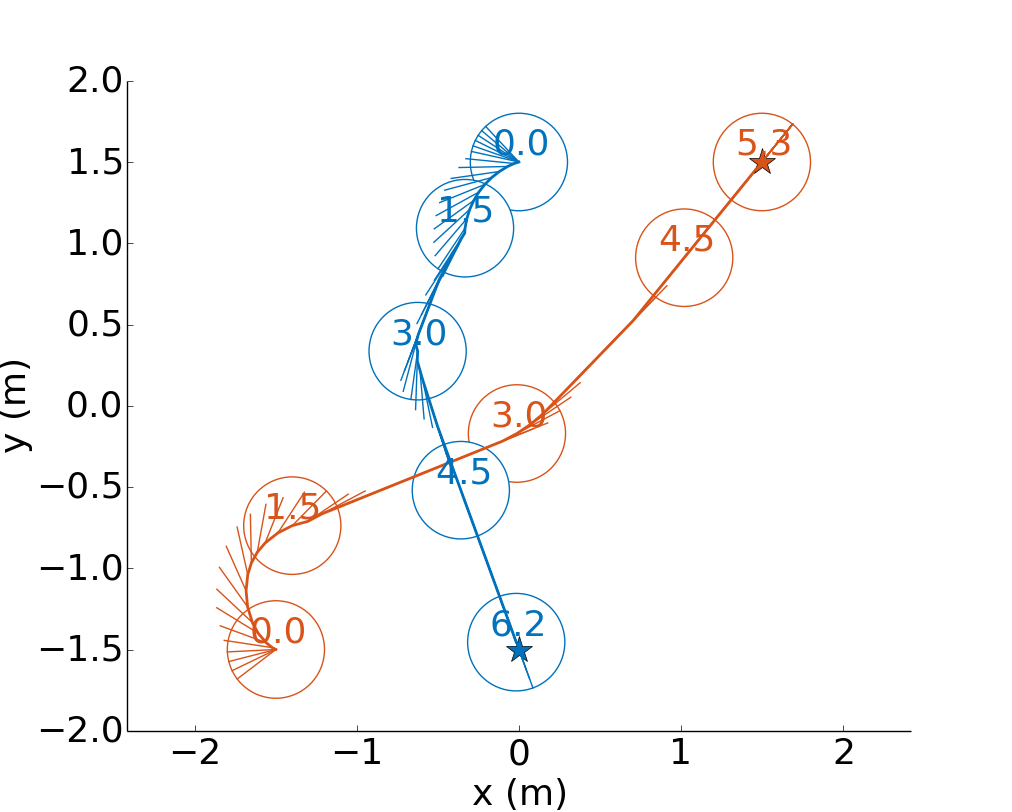}
		\caption{Sample trajectories}
		\label{fig:rotate_constr_b} 
	\end{subfigure}
	\caption{Rotational kinematic constraint. Top left shows the same test case as in \cref{fig:input_output_no_constr_a}, but here the red agent chooses to slow down due to a rotational kinematic constraint~\cref{eqn:rot_constr_1}. Bottom left shows the set of permissible velocity vectors for the red agent. Right shows a pair of sample trajectories generated by CADRL with rotational constraints, where the thin lines show the agents' heading angles. To minimize the time to goal, the red agent initially turns on the spot, and then starts moving (while continuing to turn) before its heading angle is aligned with its goal.}
	\label{fig:rotate_constr} 
	\vskip -0.1in
\end{figure}

Kinematics constraints need to be considered for operating physical robots. Yet, in many existing works, such constraints can be difficult to encode and might lead to a substantial increase in computational complexity~\cite{van_den_berg_reciprocal_2008,augugliaro_generation_2012}. In contrast, it is straightforward to incorporate kinematic constraints in the RL framework. We impose rotational constraints,
\begin{align}
	&\mathbf{a}(\mathbf{s}) = [v_s, \; \phi ] \quad \text{for} \quad v_s < v_{pref},
			  \; \left| \phi - \theta \right | < \pi / 6 \label{eqn:rot_constr_1}\\
	&\left| \theta_{t+1} - \theta_t \right | < {\Delta t} \cdot {v_{pref}}, \label{eqn:rot_constr_2}
\end{align}
where \cref{eqn:rot_constr_1} limits the direction that an agent can travel, and \cref{eqn:rot_constr_2} specifies a maximum turning rate that corresponds to a minimum turning radius of 1.0m. \Cref{fig:rotate_constr_a} illustrates the application of these rotational constraints to the same test case in \cref{fig:input_output_no_constr_a}. Here, the red agent chooses to slow down given the set of more constrained actions. Notice the agent is allowed to spin on the spot, which leads to an interesting optimal control problem when an agent's current heading angle is not perfectly aligned with its goal. In this case, an agent can either travel at its full speed while turning toward its goal, or first spin on the spot before traveling in a straight line. \Cref{fig:rotate_constr_b} shows that CADRL has learned a policy that balances between these two options to minimize the time to goal. With the thin lines showing its heading angle, the red agent chooses to initially turn on the spot, and then start moving before its heading angle is perfectly aligned with its goal. 

\subsection{Multiagent Collision Avoidance}
\begin{figure}[t]
	\centering
	\begin{subfigure}{0.23\textwidth}
	\centering
	\includegraphics [trim=10 10 0 0, clip, width=1.0 \textwidth, angle = 0]{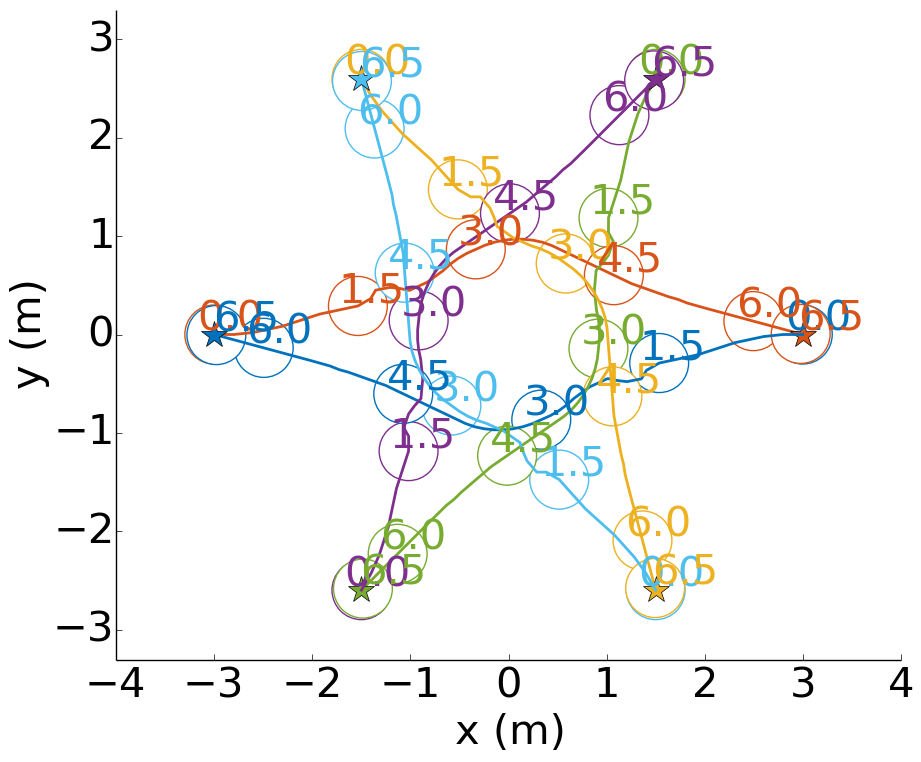}
	\caption{}
	\label{fig:multi_traj_a} 
	\end{subfigure}
	\begin{subfigure}{0.23\textwidth}
	\centering
	\includegraphics [trim=10 10 0 0, clip, width=1.0 \textwidth, angle = 0]{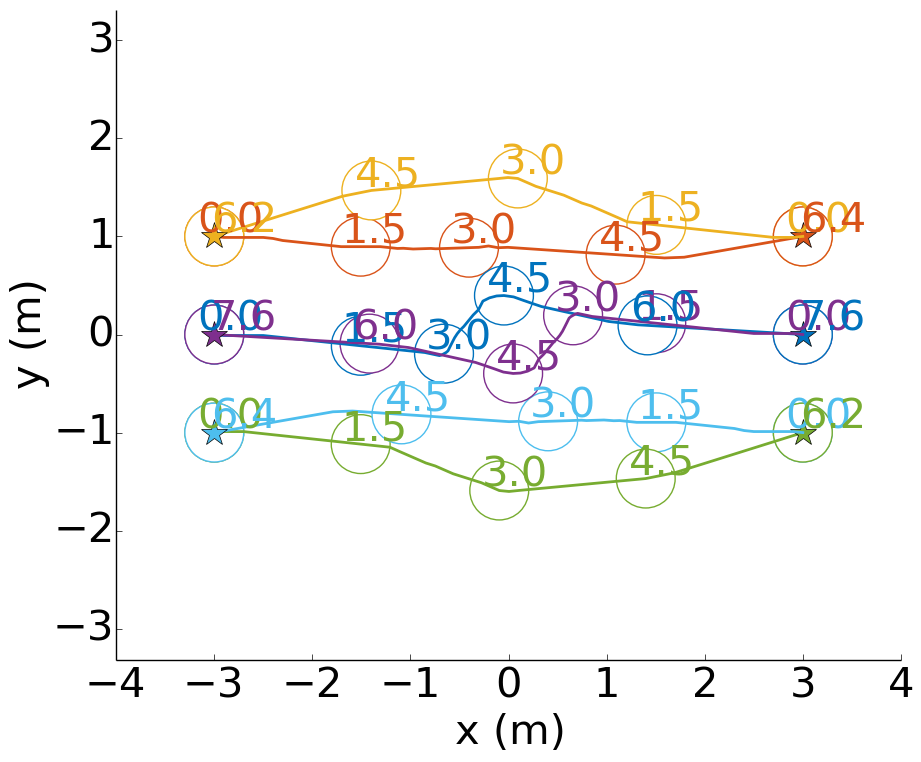}
	\caption{}
	\label{fig:multi_traj_b} 
	\end{subfigure}
	\caption{Multiagent trajectories produced by CADRL. Circles show each agent's position at the labeled time, and stars mark the goals. Although CADRL uses a two-agent value network for multiagent scenarios \cref{eqn:multi}, more complex interaction patterns have emerged. In particular, although both test cases involve three pairs of agents swapping position, each agent follows a path much different from the two agent case shown in the bottom row of \cref{fig:training}.}
	\label{fig:multi_traj} 
	\vskip -0.1in
\end{figure}

The two-agent value network can also be used for multiagent collision avoidance. Let $\tilde{\mathbf{s}}_i^o$ denote the observable part of the $i$th neighbor's state, and $\mathbf{s}^{jn}_i = [\mathbf{s}, \, \tilde{\mathbf{s}}_i^o]$ denote the joint state with the $i$th neighbor. CADRL (\cref{alg:CADRL}) can be extended to $n>2$ agents by propagating every neighbor's state one step forward (line 5-6), and then selecting the action that has the highest value with respect to any neighbor's projected state; that is, replace line 8 with  
\begin{align}
\argmax_{\mathbf{a}_t \in A} \; \min_i \; R(\mathbf{s}^{jn}_{i,t}, \; \mathbf{a}_t) + \gamma^{\Delta t \cdot v_{pref}} V(\hat{\mathbf{s}}_{t+1}, \; \hat{\tilde{\mathbf{s}}}^o_{i,t+1}). \label{eqn:multi}
\end{align}  
Note that the agent's projected next state $\hat{\mathbf{s}}_{t+1}$ also depends on the selected action $\mathbf{a}_t$.
Although using a two-agent value network, CADRL can produce multiagent trajectories that exhibit complex interaction patterns. \Cref{fig:multi_traj_a} shows six agents each moving to the opposite side of a circle. The agents veer more than the two-agent case (bottom row of \cref{fig:training}), which makes more room in the center to allow every agent to pass smoothly. \Cref{fig:multi_traj_b} shows three pairs of agents swapping position horizontally. The pair in the center slows down near the origin so the outer agents can pass first. Both cases demonstrate that CADRL can produce smooth, natural looking trajectories for multiagent systems, which will be explored in further detail in \cref{sec:results}. However, we acknowledge that \cref{eqn:multi} is only an approximation to a true multiagent RL value function -- an $n$-agent value network by simulating $n$ agents navigating around each other -- which will be studied for future work.

\section{Results} \label{sec:results}

\subsection{Computational Complexity}
This work uses a neural network with three hidden layers of width $(150, \, 100, \, 100)$, which is the size chosen to achieve real-time performance.\footnote{We also experimented with other network structures. For example, a network with three hidden layers of width $(300, \, 200, \, 200)$ produced similar results (paths) but was twice as slow.} In particular, on a computer with an i7-5820K CPU, a Python implementation of CADRL (\cref{alg:CADRL}), on average, takes 5.7ms per iteration on two-agent collision avoidance problems. By inspection of~\cref{eqn:multi}, computational time scales linearly in the number of neighboring agents for a decentralized implementation where each agent runs CADRL individually; and scales quadratically in a centralized implementation where one computer controls all agents. For decentralized control on ten agents, each iteration of CADRL is observed to take 62ms. Moreover, CADRL is parallelizable because it consists of a large number of independent queries of the value network \cref{eqn:multi}. 

Furthermore, offline training (\cref{alg:V-learning}) took less than three hours and is found to be quite robust. In particular, using mini-batches of size 500, the initialization step (line 3) took 9.6 minutes to complete 10,000 iterations of back-propagation. The RL step used an $\epsilon$-greedy policy, where $\epsilon$ decays linearly from 0.5 to~0.1 in the first 400 training episodes, and remains~0.1 thereafter. The RL step took approximately~2.5 hours to complete 1,000 training episodes. The entire training process was repeated on three sets of training trajectories generated by ORCA, and the value network converged in all three trials. The paths generated by the value networks from the three trials were very similar, as indicated by a less than~5\% difference in time to reach goal on all test cases in the evaluation set.

\subsection{Performance Comparison on a Crossing Scenario}
To evaluate the performance of CADRL over a variety of test cases, we compute the average extra time spent to reach goals, that is
\begin{align}
\bar{t}_e = \frac{1}{n} \sum_{i=1}^n \left[ t_{i,g} - \frac{||\mathbf{p}_{i,0} - \mathbf{p}_{i,g}||_2}{v_{i, pref}} \right], \label{eqn:extra_time}
\end{align}
where $t_{i,g}$ is the $i$th agent's time to reach its goal, and the second term is a lower bound of $t_{i,g}$ (to go straight toward goal at the preferred speed). This metric removes the effects due to variability in the number of agents and the nominal distance to goals.

\begin{figure}[t]
	\centering
	\begin{subfigure}{0.22\textwidth}
	\centering
	\includegraphics [trim=50 60 550 130, clip, height=0.9 \textwidth, angle = 0]{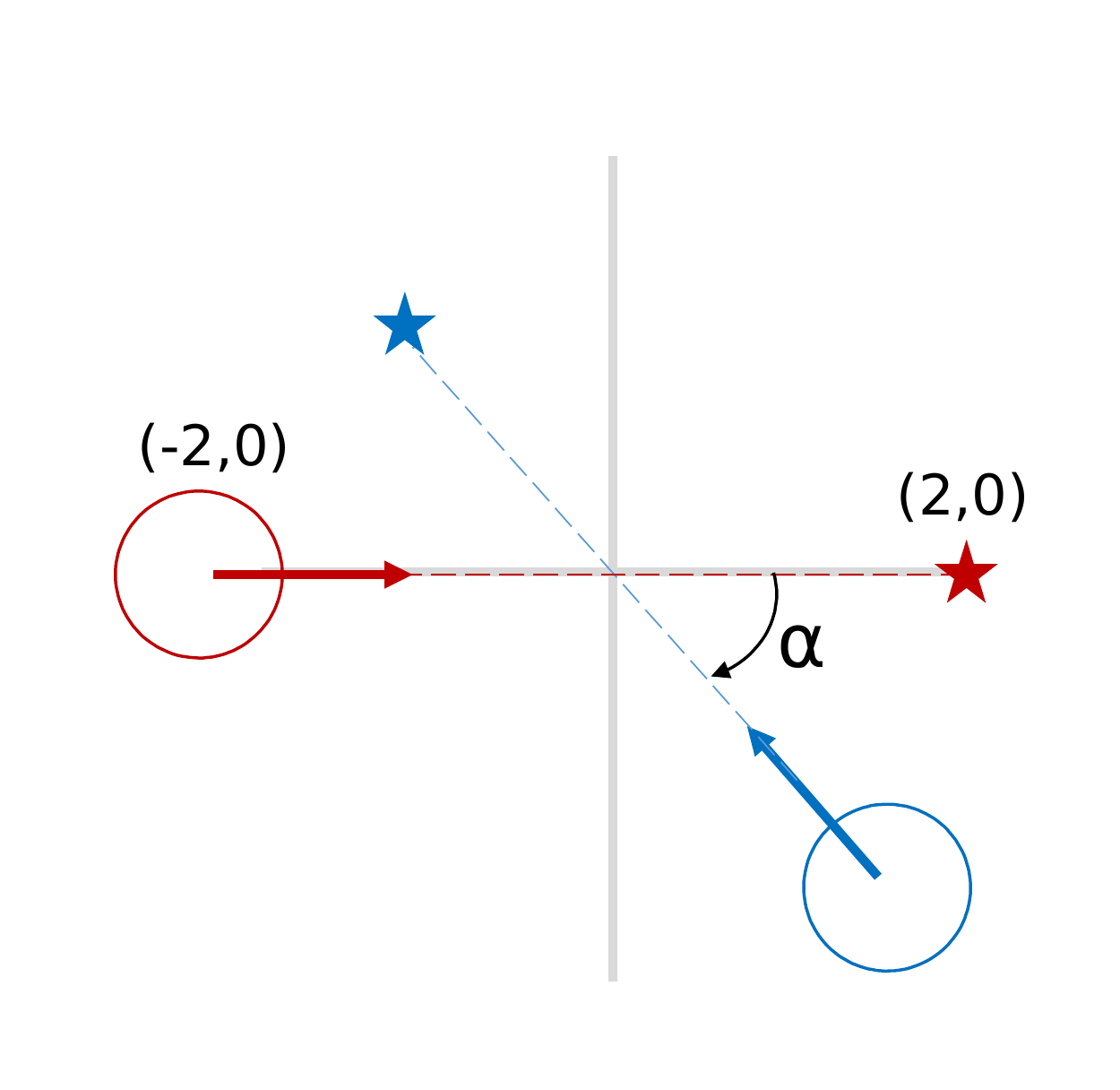}
	\caption{Crossing configuration}
	\label{fig:crossing_a} 
	\end{subfigure}
	\begin{subfigure}{0.22\textwidth}
	\centering
	\includegraphics [trim=0 0 0 0, clip, height=0.9 \textwidth, angle = 0]{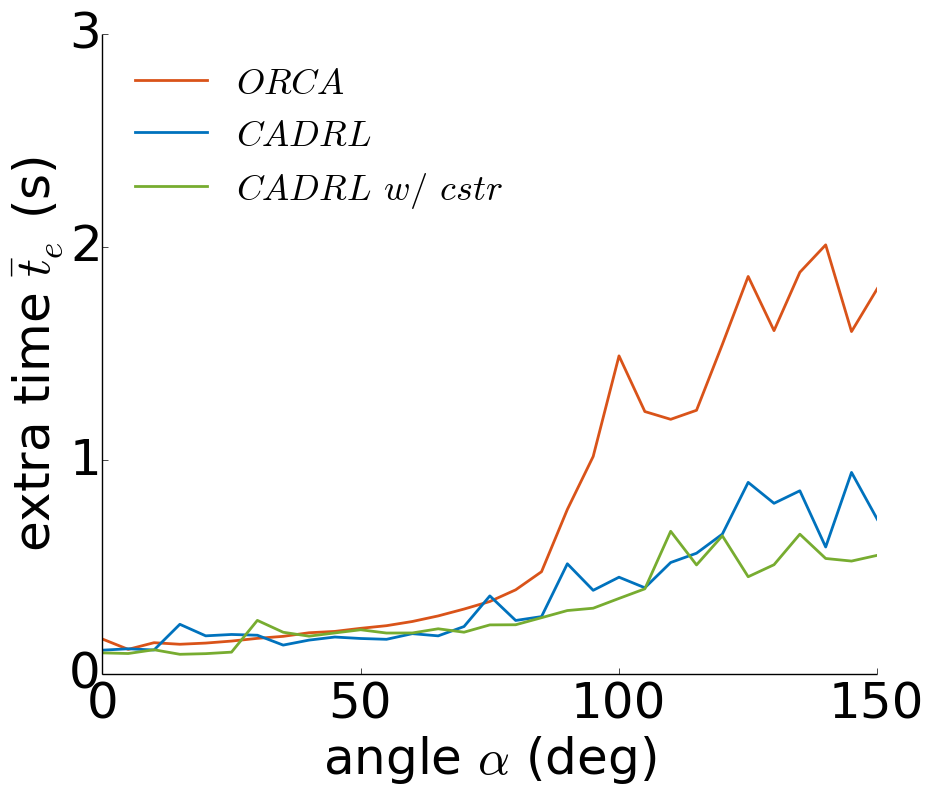}
	\caption{Time to reach goal}
	\label{fig:crossing_b} 
	\end{subfigure}
	\caption{Performance comparison on a crossing scenario. (a) shows the crossing configuration, where a red agent travels from left to right, and a blue agent travels at a diagonal angled at $\alpha$. Both agents have a radius of 0.3m and a preferred speed of 1.0m/s, and they would collide at the origin if both travel in a straight line. (b) compares the extra time spent to reach goal \cref{eqn:extra_time} using different collision avoidance strategies. CADRL performs significantly better than ORCA on a wide range of angles $\alpha \in [90, 150]$ deg.} 
	\label{fig:crossing} 
	\vskip -0.1in
\end{figure}

A crossing scenario is shown in \cref{fig:crossing_a}, where two identical agents with goals along collision courses are run into each other at different angles. Thus, cooperation is required for avoiding collision at the origin. \Cref{fig:crossing_a} shows that over a wide range of angles ($\alpha \in [90, 150]$ deg), agents following CADRL reached their goals much faster than that of ORCA. Recall a minimum separation of 0.2m is specified for CADRL in the reward function~\cref{eqn:reward}. For this reason, similar to the bottom row of \cref{fig:training}, CADRL finds paths that are slightly slower than ORCA around $\alpha=0$. It is also interesting to note that CADRL with rotational constraints (CADRL w/ cstr) performs slightly better than the unconstrained. This is because CADRL w/ cstr is more conservative (yielding) early on, which is coincidentally good for the crossing scenario. More specifically, if the other agent has stopped (reached goal) or turned before it reached the origin, unconstrained CADRL would have performed better. In short, as will be shown later in \cref{fig:scatter}, unconstrained CADRL is better on average (randomized test cases), but can be slightly worse than CADRL w/ cstr on particular test cases.  

\subsection{Performance Comparison on Random Test Cases}
\begin{figure}[t]
	\centering
	\begin{subfigure}{0.22\textwidth}
		\centering
		\includegraphics [trim=0 0 0 0, clip, width=1.0 \textwidth, angle = 0]{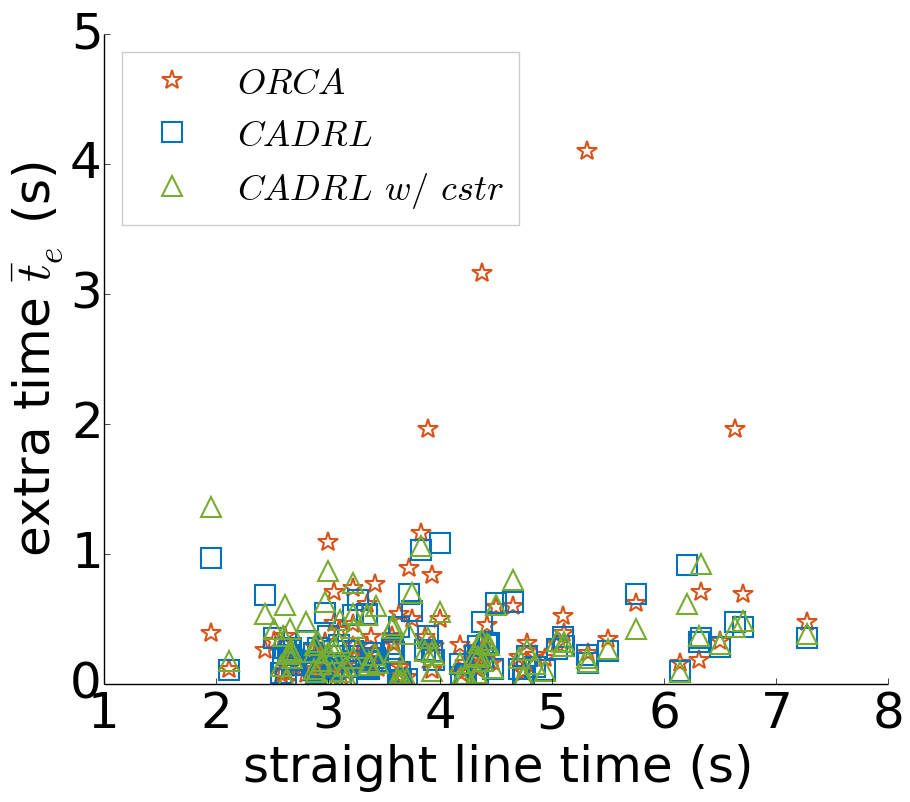}
		\caption{2 agents}
		\label{fig:scatter_a} 
	\end{subfigure}
	\begin{subfigure}{0.22\textwidth}
		\centering
		\includegraphics [trim=0 0 0 0, clip, width=1.0 \textwidth, angle = 0]{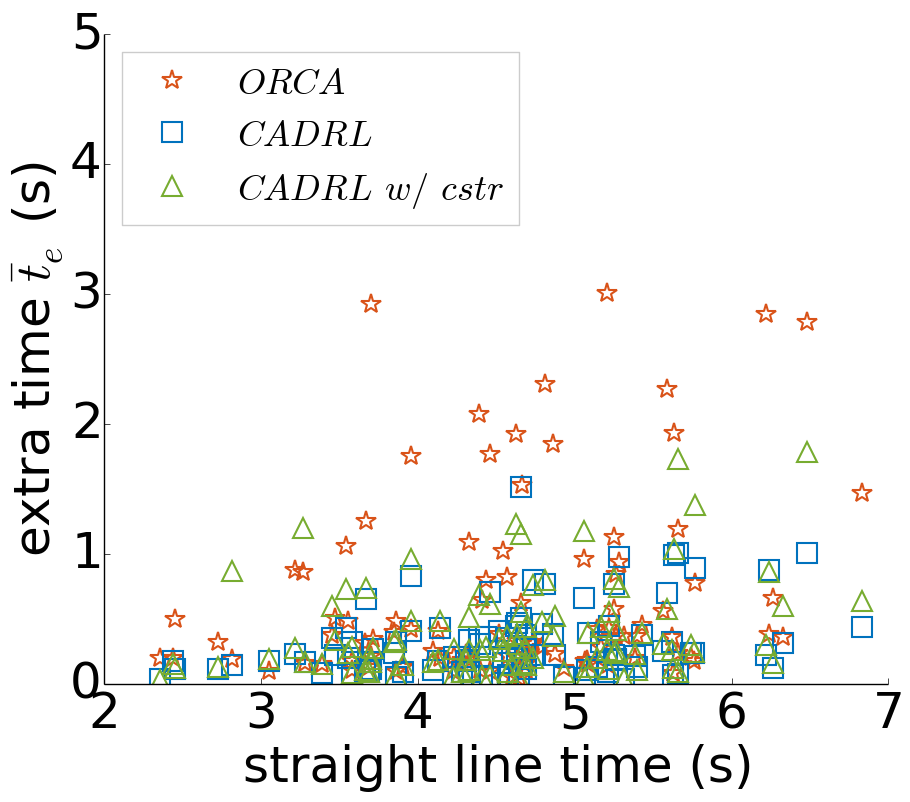}
		\caption{4 agents}
		\label{fig:scatter_b} 
	\end{subfigure}
	
	\begin{subfigure}{0.46\textwidth}
	\centering
	\includegraphics [trim=0 0 0 0, clip, width=0.9 \textwidth, angle = 0]{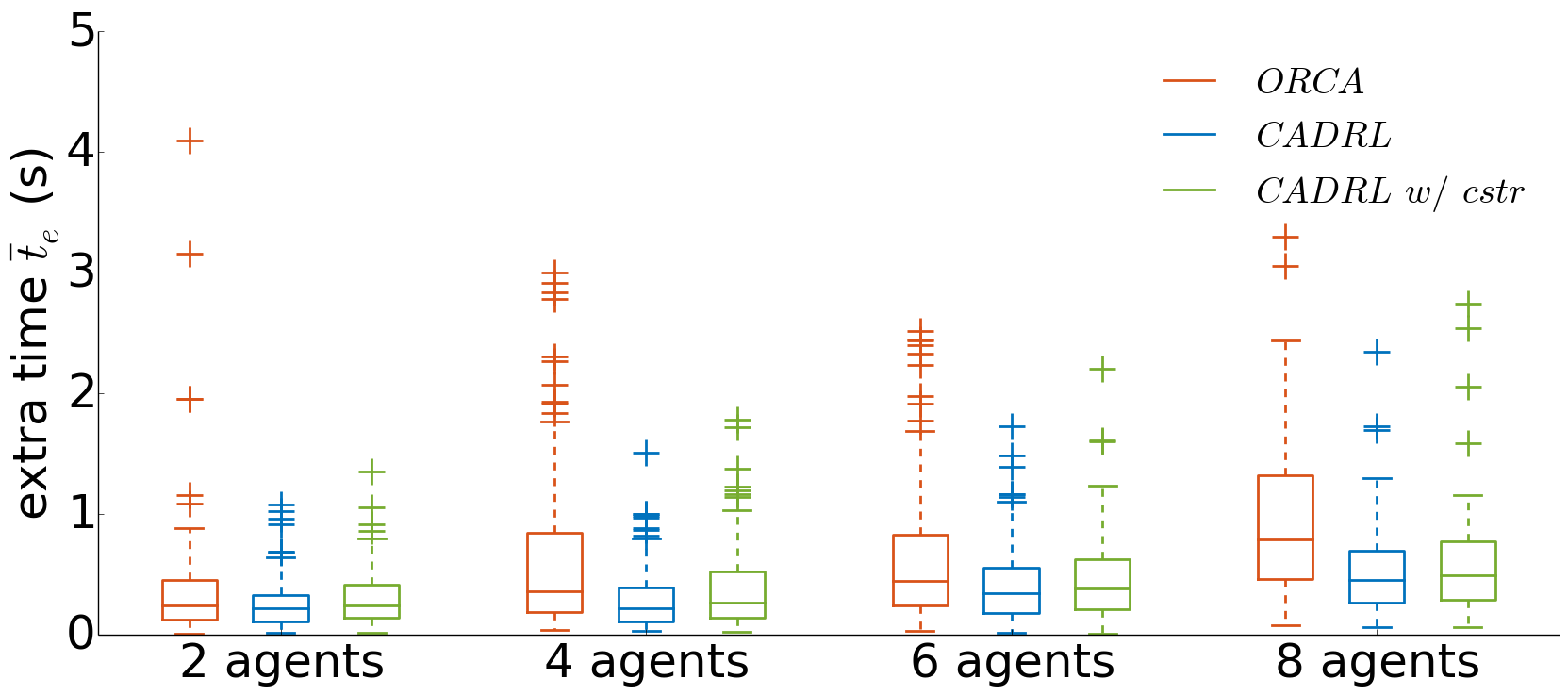}
	\caption{Box-and-whisker plot}
	\label{fig:scatter_c} 
	\end{subfigure}
	\caption{Performance comparison on randomly generated test cases. The extra time to goal $t_e$ is computed on one hundred random test cases for each configuration listed in \cref{tab:stat}. (a) and (b) show scatter plots of the raw data, and (c) shows a box whisker plot. CADRL is seen to perform similarly (slightly better) to ORCA on the easier test cases (median), and significantly better on the more difficult test cases ($>$75 percentile).}
	\label{fig:scatter} 
	\vskip -0.1in
\end{figure}

In addition to showing that CADRL can handle some difficult test cases that fared poorly for ORCA (\cref{fig:training,fig:crossing}), a more thorough evaluation is performed on randomly generated test cases. In particular, within square shaped domains specified in \cref{tab:stat}, agents are generated with randomly sampled speed, radius, initial positions and goal positions. This work chooses $v_{pref} \in [0.5, 1.5]$m/s, $r \in [0.3, 0.5]$m, which are parameters similar to that of typical pedestrian motion. Also, the agents' goals are projected to the boundary of the room to avoid accidentally creating a trap formed by multiple stationary agents. A sample four-agent test case is illustrated in \cref{fig:multi_traj}, where agents following CADRL were able to reach their goal much faster than that of ORCA. For each configuration in \cref{tab:stat}, one hundred test cases are generated as described above. ORCA, CADRL, and CADRL w/cstr are employed to solve for these test cases. The average extra time to reach goal, $\bar{t}_e$, is computed for each set of generated trajectories and plotted in \cref{fig:scatter}. Key statistics are computed and listed in \cref{tab:stat}, and it can be seen that CADRL performs similarly (slightly better) than ORCA on the easier test cases (median), and more than 26\% better on the hard test cases ($>$75 percentile). Also, performance difference is more clear on test cases with more agents, which could be a result of more frequent interactions.

\begin{figure}[t]
	\centering
	\begin{subfigure}{0.23\textwidth}
	\centering
	\includegraphics [trim=10 10 0 0, clip, width=1.0 \textwidth, angle = 0]{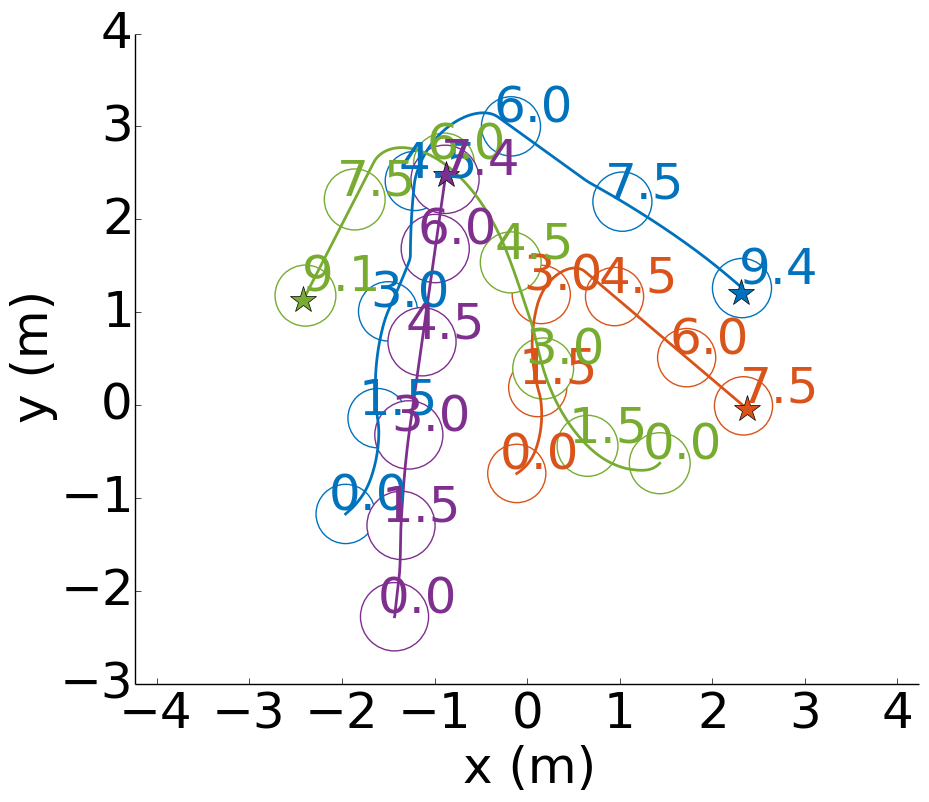}
	\caption{ORCA}
	\end{subfigure}
	\begin{subfigure}{0.23\textwidth}
	\centering
	\includegraphics [trim=10 10 0 0, clip, width=1.0 \textwidth, angle = 0]{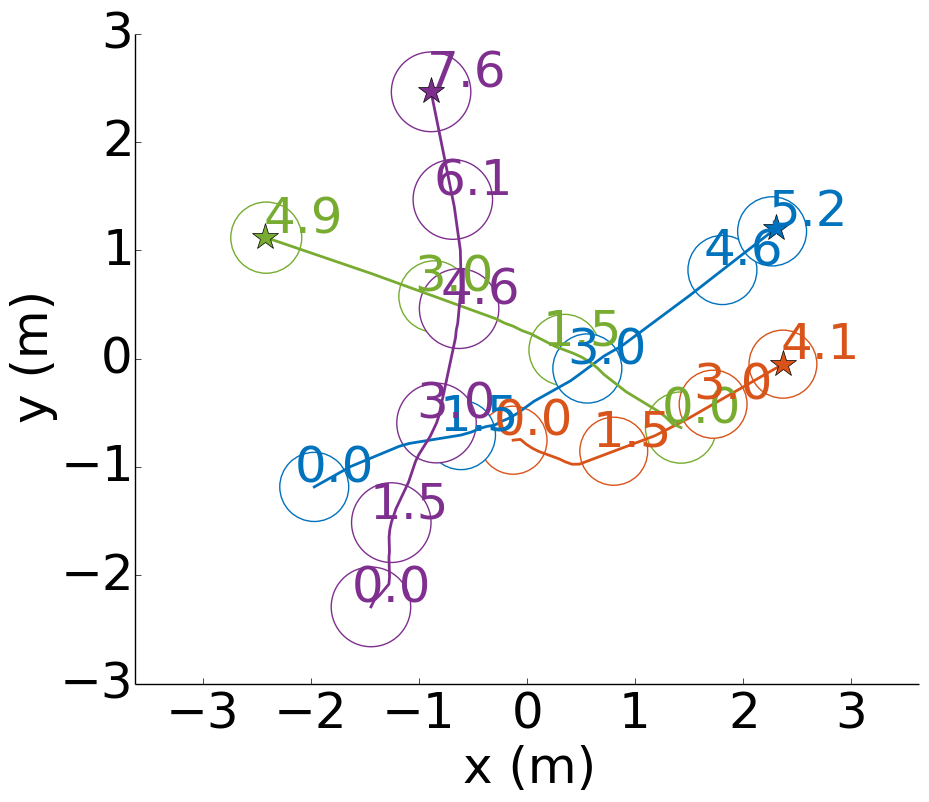}
	\caption{CADRL}
	\end{subfigure}
	\caption{Four-agent test case. (a) shows agents following ORCA traversed long arcs before reaching their goals, which reflects the similar two-agent problem shown in \cref{fig:training_a}. (b) shows agents following CADRL were able to reach their goal much faster. }
	\label{fig:multi_traj} 
	\vskip -0.1in
\end{figure}

\begin{table*}[t]
	\centering
	\caption[]{Extra time to reach goal ($\bar{t}_e$) statistics on the random test cases shown in \cref{fig:scatter}. CADRL finds paths that on average, reach goals much faster than that of ORCA. The improvement is more clear on hard test cases ($>$75 percentile) and in multiagent ($n>2$) scenarios that require more interactions.} 
	\begin{tabular}{|c | c || c | c | c || c | c | c |}
	\hline
		\multicolumn{2}{|c||}{\rule{0pt}{8pt} Test case configuration}  & \multicolumn{3}{c||}{Extra time to goal $\bar{t}_e$ (s) [Avg / 75th / 90th percentile]} & \multicolumn{3}{c|}{Average min separation dist. (m)} 
		\\
		\cline{1-8}
		 \rule{0pt}{8pt} {num agents} & {domain size (m)} & ORCA & CADRL & CADRL w/ cstr & OCRA & CADRL & CADRL w/ cstr \\ \hline
		\rule{0pt}{8pt} {2} &  4.0 $\times$ 4.0 & 0.46 / 0.45 / 0.73 & 0.27 / 0.33 / 0.56 &   0.31 / 0.42 / 0.60 &  0.122 & 0.199 & 0.198 \\
		\rule{0pt}{8pt} {4} &  5.0 $\times$ 5.0 & 0.69 / 0.85 / 1.85 & 0.31 / 0.40 / 0.76 & 0.39 / 0.53 / 0.86 & 0.120 & 0.192 & 0.191 \\
		\rule{0pt}{8pt} {6} & 6.0 $\times$ 6.0 & 0.65 / 0.83 / 1.50 &  0.44 / 0.56 / 0.87 & 0.48 / 0.63 / 1.02 & 0.118 & 0.117 & 0.180 \\
		\rule{0pt}{8pt} {8} & 7.0 $\times$ 7.0 & 0.96 / 1.33 / 1.84 & 0.54 / 0.70 / 1.01 & 0.59 / 0.77 / 1.09 & 0.110 & 0.171 & 0.170 \\ \hline
	\end{tabular} 
	\label{tab:stat}
	\vskip -0.1in
\end{table*}

\subsection{Navigating around Non-cooperative Agents} \label{sec:results:non_coop}
\begin{figure}[t]
	\centering
	\begin{subfigure}{0.23\textwidth}
	\centering
	\includegraphics [trim=0 0 0 35, clip, width=1.0 \textwidth, angle = 0]{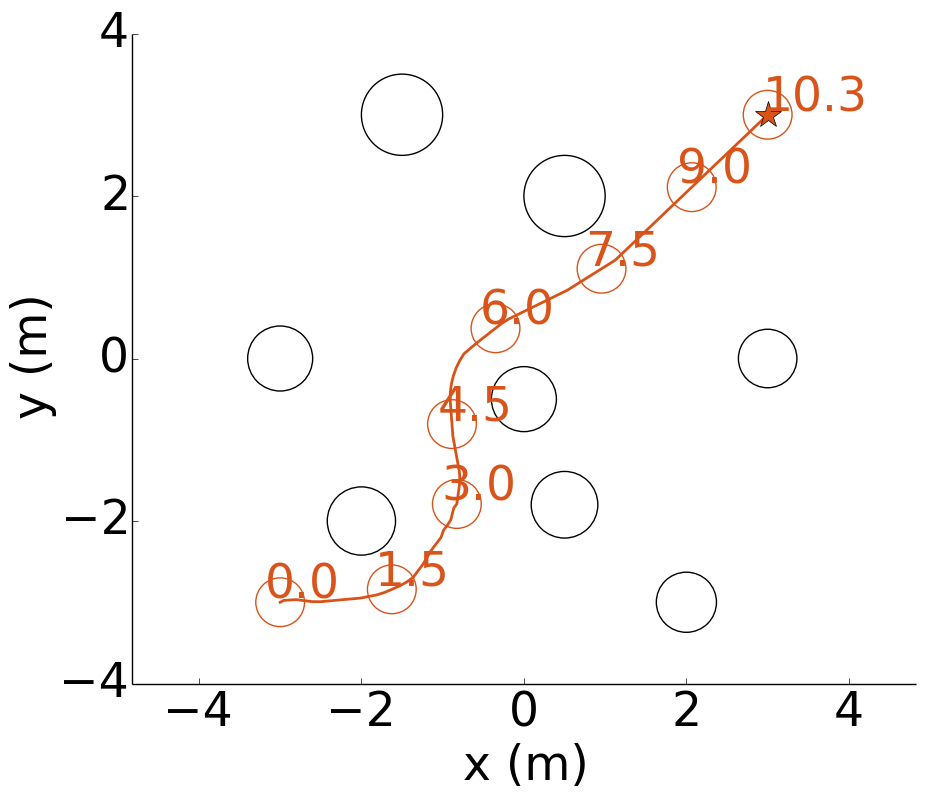}
	\caption{Static obstacles}
	\label{fig:non_coop_a} 
	\end{subfigure}
	\begin{subfigure}{0.23\textwidth}
	\centering
	\includegraphics [trim=0 0 0 10, clip, width=1.0 \textwidth, angle = 0]{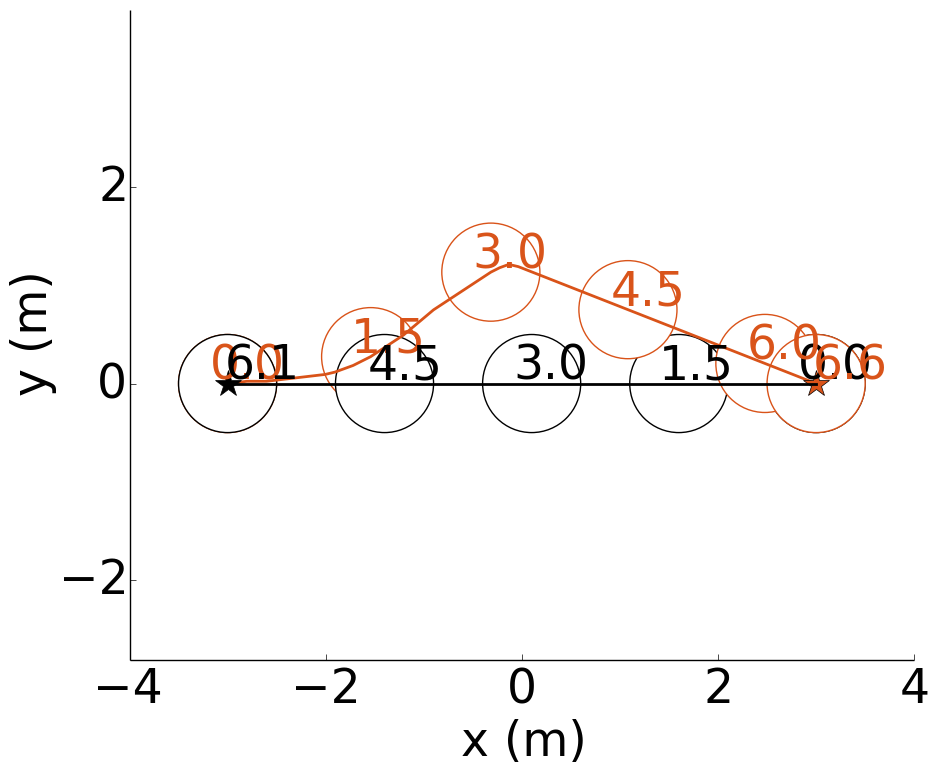}
	\caption{Non-cooperative agent}
	\label{fig:non_coop_b} 
	\end{subfigure}
	\caption{Navigating around non-CADRL agents. (a) shows a red agent navigating around a series of static obstacles. (b) shows a red CADRL agent avoids collision with a non-cooperative black agent, who traveled in a straight line from right to left.}
	\label{fig:non_coop} 
	\vskip -0.2in
\end{figure}

Recall CADRL's value network is trained with both agents following the same collision avoidance strategy, which is the reciprocity assumption common to many existing works ~\cite{van_den_berg_reciprocal_2008,berg_reciprocal_2011}. \Cref{fig:non_coop} shows that the proposed method can also navigate efficiently around non-CADRL agents. \Cref{fig:non_coop_a} shows a CADRL agent navigating static obstacles modeled as stationary agents ($\tilde{\mathbf{v}}_i = 0$). We acknowledge that CADRL could get stuck in a dense obstacle field, where traps/dead-ends could form due to positioning of multiple obstacles. Recall CADRL is a collision avoidance (not path planning) algorithm not designed for such scenarios. \Cref{fig:non_coop_b} shows a CADRL agent navigating around a non-cooperative agent (black), who traveled in a straight line from right to left. In comparison with the cooperative case shown in \cref{fig:training_d}, here the red CADRL agent chooses to veer more to its left to avoid collision.   

\section{Conclusion} \label{sec:conclusion}
This work developed a decentralized multiagent collision avoidance algorithm based on a novel application of deep reinforcement learning. In particular, a pair of agents were simulated to navigate around each other to learn a value network that encodes the expected time to goal. The solution (value network) to the two-agent collision avoidance RL problem was generalized in a principled way to handle multiagent ($n>2$) scenarios. The proposed method was shown to be real-time implementable for a decentralized ten-agent system. Simulation results show more than 26\% improvement in paths quality when compared with ORCA. 





\section*{Acknowledgment}
This work is supported by Ford Motor Company.
 
\bibliographystyle{IEEEtran} 
\bibliography{biblio}
\balance
\end{document}